\PassOptionsToPackage{table,xcdraw}{xcolor}

\PassOptionsToPackage{prologue,dvipsnames}{xcolor}
\documentclass[sigconf,screen]{acmart}

\usepackage[table,xcdraw]{xcolor}
\usepackage{tikz}
\usetikzlibrary{patterns}
\usepackage{multirow}
\usepackage[table,xcdraw]{xcolor}
\usepackage{colortbl}
\usepackage{tabularx}
\usepackage{graphicx}
\AtBeginDocument{%
  }

\copyrightyear{2025}
\acmYear{2025}
\setcopyright{acmlicensed}\acmConference[CHI '25]{CHI Conference on Human Factors in Computing Systems}{April 26-May 1, 2025}{Yokohama, Japan}
\acmBooktitle{CHI Conference on Human Factors in Computing Systems (CHI '25), April 26-May 1, 2025, Yokohama, Japan}
\acmDOI{10.1145/3706598.3714120}
\acmISBN{979-8-4007-1394-1/2025/04}

\begin{document}

\title{Understanding Screenwriters' Practices, Attitudes, and Future Expectations in Human-AI Co-Creation}


\author{Yuying Tang}
\affiliation{%
  \institution{The Hong Kong University of Science and Technology }
  \city{Hong Kong SAR}
  \country{China}
}
\email{ytangdh@connect.ust.hk}

\author{Haotian Li}
\authornote{The work was done when Haotian Li was at HKUST. Haotian Li is the corresponding author.}
\affiliation{%
  \institution{Microsoft Research Asia}
  \city{Beijing}
  \country{China}
}
\email{haotian.li@microsoft.com}

\author{Minghe Lan}
\affiliation{%
  \institution{Central Academy of Fine Arts}
  \city{Beijing}
  \country{China}
}
\email{lan_shan99@outlook.com}

\author{Xiaojuan Ma}
\affiliation{%
  \institution{The Hong Kong University of Science and Technology}
  \city{Hong Kong SAR}
  \country{China}
}
\email{mxj@cse.ust.hk}

\author{Huamin Qu}
\affiliation{%
  \institution{The Hong Kong University of Science and Technology}
  \city{Hong Kong SAR}
  \country{China}
}
\email{huamin@cse.ust.hk}

\newcommand{\haotian}[1]{\textcolor{teal}{#1}}
\newcommand{\revision}[1]{\textcolor{red}{#1}}



\begin{abstract}


With the rise of AI technologies and their growing influence in the screenwriting field, understanding the opportunities and concerns related to AI's role in screenwriting is essential for enhancing human-AI co-creation. Through semi-structured interviews with 23 screenwriters, we explored their creative practices, attitudes, and expectations in collaborating with AI for screenwriting. Based on participants' responses, we identified the key stages in which they commonly integrated AI, including story structure \& plot development, screenplay text, goal \& idea generation, and dialogue. Then, we examined how different attitudes toward AI integration influence screenwriters' practices across various workflow stages and their broader impact on the industry. Additionally, we categorized their expected assistance using four distinct roles of AI: actor, audience, expert, and executor. Our findings provide insights into AI's impact on screenwriting practices and offer suggestions on how AI can benefit the future of screenwriting.

\end{abstract}

\maketitle

\section{Introduction}

\textcolor{black}{Screenwriting involves the creation of screenplays that integrate visual and auditory elements to support storytelling in film, television, and other audiovisual formats~\cite{dunnigan2019screenwriting}, with the aim of evoking emotional responses through detailed plot structures, dialogue, and scenes~\cite{howard1993tools}. 
Belonging to the area of creative writing, screenwriting also has an imaginative and narrative focus~\cite{kerrigan2016re, batty2015screenwriter}.
As a result, it faces similar challenges to creative writing, such as a lack of inspiration and guidance~\cite{10.1145/3544548.3580782}. 
Futhermore, its distinctive emphasis on incorporating visual and auditory considerations adds layers of complexity to the creative processes and necessitates tailored approaches to addressing these challenges~\cite{duncan2020guide}.}

The rapid advancement of technology has enabled the application of AI in screenwriting~\cite{anguiano2023hollywood}. Previous research demonstrates that AI can support screenwriters by summarizing and cross-checking content to ensure the consistency of screenplay text~\cite{sanghrajka2017lisa, kapadia2015computer}, as well as analyzing emotions through plot units~\cite{goyal2010toward}. \textcolor{black}{With developments ranging from natural language processing (NLP) to deep learning (DL), AI capabilities have evolved from processing and summarizing content to generating it, offering innovative solutions to screenwriting challenges.} For instance, AI can generate character roles with diverse appearances and dialogue styles~\cite{10.1145/3613904.3642105}, assisting users in sparking creative ideas.

\textcolor{black}{These advancements in AI have shown their ability to enhance screenwriters’ productivity and creativity, leading to economic and industry-wide impacts~\cite{chow2020ghost}. The role of AI in screenwriting was notably highlighted during the Hollywood writers' strike when the Writers Guild of America and production companies reached an agreement defining AI as a supporting tool rather than a replacement for writers~\cite{nyt_wga_strike_ai}. The agreement prohibits AI from receiving credit, ensures fair compensation for human writers, and allows optional AI assistance for tasks such as script drafting. This reflects the growing integration of advanced AI technologies in screenwriting and their transformative impact on the industry, motivating further exploration of AI’s role in screenwriting.}

\textcolor{black}{While prior studies highlight AI’s expanding role and its impact on screenwriting education~\cite{brako2023robots} and the broader filmmaking process~\cite{naji2024employing}, there remains limited understanding of how AI is applied across different stages of the screenwriting workflow. To address this gap, we propose \textbf{RQ1}, which seeks to explore screenwriters’ current practices, including their task allocation involving AI tools.}

\textcolor{black}{Additionally, existing research has primarily focused on topics such as attitudes~\cite{10.1145/3656650.3656688}, copyright~\cite{kavitha2023copyright}, and labor considerations~\cite{chow2020ghost}. However, it provides limited insights into the nuanced relationship between screenwriters’ diverse attitudes toward AI and its practical effects across various stages of their workflows. To address this gap, we propose \textbf{RQ2}, which focuses on understanding the reasons behind screenwriters’ differing attitudes and how these attitudes influence their practices regarding AI integration and their perceptions of AI’s impact.}

\textcolor{black}{Finally, recent work by Grigis and Angeli examined the features, strategies, and outcomes of LLM-assisted playwriting~\cite{10.1145/3656650.3656688}, and the Dramatron system, developed by Mirowski et al.~\cite{10.1145/3544548.3581225}, demonstrated the potential of integrating LLMs into screenwriting workflows. However, these studies focus solely on LLMs, neglecting advancements in AI-generated images and videos. A broader understanding of how emerging and rapidly evolving AI technologies may address the dynamic and evolving needs of screenwriters is still lacking. To address this gap, we propose \textbf{RQ3}, which aims to explore screenwriters’ detailed expectations for future AI tools.}

\begin{enumerate}
    \item[\textbf{RQ1:}] How is AI involved in existing practices within the screenwriting workflow?
    \item[\textbf{RQ2:}] What are the attitudes of screenwriters towards the integration of AI into their workflow?
    \item[\textbf{RQ3:}] How do screenwriters envision the future roles of AI in the creative process?
\end{enumerate}











We conducted in-depth interviews with 23 screenwriters to address these research questions. For \textbf{RQ1}, most participants (78\%) reported integrating AI into workflows, primarily in stages such as story structure \& plot, screenplay text, goal \& idea generation, and dialogue. \textcolor{black}{For \textbf{RQ2}, participants highlighted AI's strengths in reducing trial-and-error costs, expanding knowledge boundaries, and inspiring overlooked creative ideas but noted challenges such as high usage barriers, inaccuracy, and limited emotional understanding. Opinions on structured and pastiche generation were mixed. Regarding broader societal impacts, participants expressed optimism about AI’s potential to support divergent thinking and industry communication but raised concerns about authorship and copyright. Views on AI as a competitor in screenwriting were divided.} For \textbf{RQ3}, we identified participants' desired AI features in four roles: ``actor'' for character simulation, ``audience'' for feedback, ``expert'' for professional advice, and ``executor'' for completing tasks. These findings provide insights to guide the development of future AI tools for screenwriting.

In summary, our study offers the following contributions:

\begin{itemize}
\item We provide insights into the current AI-involved screenwriting practices, revealing the multilayered and diverse ways in which AI tools are utilized.

\item We summarize the complex attitudes screenwriters hold towards AI technology and identify the specific stages within the workflow where AI will be needed in the future.

\item \textcolor{black}{We identify four potential roles that AI could play in screenwriting, aiming to clarify screenwriters' specific functional expectations for AI tools in future co-creation.}
\end{itemize}

\section{Related Work}
\textcolor{black}{Our study builds on prior research on AI integration in creative processes, AI tools for creative writing, and AI-supported screenwriting, highlighting gaps specific to screenwriting. These gaps stem from its unique characteristics, including audiovisual storytelling~\cite{senje2017formatting}, structured principles~\cite{mckee1997substance}, and multi-stakeholder collaboration~\cite{taylor2024one, cake2021collaborative, batty2018script}. Previous studies have not fully incorporated evolving AI capabilities to consider these screenwriting characteristics, underscoring the need for further investigation.}

\subsection{\textcolor{black}{AI Integration in Creative Processes}}

\textcolor{black}{Advancements in AI technology have significantly impacted various creative domains~\cite{anantrasirichai2022artificial, 10.1145/3613904.3642726}, reshaping fields such as art~\cite{cetinic2022understanding} and design~\cite{lee2023impact}. In art, AI has contributed to interactive media and visual representation~\cite{anadol2018archive, sun2023ai, tang2023ai}. In design, AI has been integrated into product design workflows, fostering innovation in idea generation~\cite{hong2023generative, chiou2023designing} and advancing user interface design through features such as adaptability~\cite{cheng2023play, zhang2023layoutdiffusion}. Furthermore, the impact of AI-generated image tools on workflows has been explored~\cite{mustafa2023impact, tang2024s}, highlighting AI’s potential to democratize graphic design~\cite{tang2024exploring} and emphasizing the need for interdisciplinary collaboration to enhance AI systems~\cite{meron2022graphic}. As AI-generated video capabilities advance, AI is also reshaping fashion design through dynamic visuals~\cite{liu2019toward, karras2023dreampose} and supporting the creation of short-form social media videos~\cite{10.1145/3613904.3642476}. Additionally, other research highlights AI's role in various creative domains, such as design~\cite{10.1145/3613904.3642812}, and data storytelling~\cite{10.1145/3613904.3642726}.}


The growing importance of effective human-AI co-creation, as highlighted in prior research within visual-based creative domains, motivates us to investigate how diverse AI capabilities and applications have impacted text-based creative work.


\subsection{\textcolor{black}{AI Tools for Creative Writing}}
\textcolor{black}{Creative writing, as a text-based creative work encompassing novels, short stories, poetry, and essays, emphasizes linguistic artistry, stylistic diversity, and imaginative narrative freedom~\cite{ramet2011creative, smith2020writing, mcvey2008all}. The development of AI technologies has significantly supported this creative field~\cite{10.1145/3613904.3642529}, evolving from natural language processing to advanced deep learning systems. This progression spans basic tools like spell checkers~\cite{peterson1980computer}, crowdsourcing platforms~\cite{kim2017mechanical, kim2014ensemble}, and crowd role-play approaches~\cite{huang2020heteroglossia}, to AI-driven character role-playing systems~\cite{10.1145/3613904.3642105}. Prior studies have shown that AI reduces creative workloads by providing valuable suggestions~\cite{gero2019stylistic, bernstein2010soylent} and refining content for professional use~\cite{hui2018introassist, roemmele2015creative, clark2018creative}, thereby helping writers overcome creative blocks. Specifically, AI-supported creative writing processes are currently categorized into story development and character creation~\cite{radford2019language, yang2019sketching, calderwood2020novelists, clark2018creative, coenen2021wordcraft}.}


\textcolor{black}{In story development, platforms such as Dramatica~\cite{dramatica} and AI Dungeon~\cite{aidungeon} support story generation. TaleStream combines AI recommendation algorithms with database-driven trope suggestions for narratives but has yet to integrate evolving generative AI technologies~\cite{chou2023talestream}. Rule-based methods for story extension have also been explored~\cite{lebowitz1984creating, meehan1977tale, riedl2010narrative}. However, we found that these approaches primarily focus on text-based support and do not incorporate visual elements. In character creation, prior works demonstrate that AI can simulate characters' personalities~\cite{cavazza2001characters}, infer characters' relationships~\cite{chaturvedi2017unsupervised}, facilitate AI-character dialogues~\cite{10.1145/3613904.3642105}, and map character trajectories~\cite{chung2022talebrush}. These tools showcase the potential of AI for inspiring character development~\cite{10.1145/3450741.3465253, shakeri2021saga, yuan2022wordcraft}, but they are often limited to isolated stages and lack integration across the creative workflow. Additionally, challenges in human-AI collaboration remain. Grigis and Angeli recently examined the limitations of LLM-assisted writing, particularly in handling taboos and conflicts~\cite{10.1145/3656650.3656688}. Dhillon et al. emphasized the importance of personalized systems to accommodate the diverse experiences of writers~\cite{10.1145/3613904.3642134}, and Lee et al. provided broad recommendations for the development of future AI tools~\cite{10.1145/3613904.3642697}. Meanwhile, Biermann et al. underscored writers’ concerns about the potential impact of AI on creativity~\cite{biermann2022tool}.}

\textcolor{black}{Based on these, we note that existing research largely focuses on text-based AI support, often overlooking the consideration of visual elements in the creative process. 
This limitation reduces the suitability of these approaches for certain types of creative writing, particularly screenwriting, which is inherently characterized by audiovisual and dynamic storytelling~\cite{senje2017formatting}. Furthermore, existing studies often focus on isolated stages of the creative process, overlooking the complete workflow that integrates structured principles~\cite{mckee1997substance} and collaboration with stakeholders~\cite{taylor2024one, cake2021collaborative, batty2018script}, both of which are essential to screenwriting. These gaps underscore the need to explore AI's current applications in screenwriting.}

\subsection{\textcolor{black}{AI-Supported Screenwriting}}

\textcolor{black}{The development of AI technologies has already influenced screenwriting~\cite{batty2015screenwriter, anguiano2023hollywood}. Previous applications of AI tools in screenwriting have primarily focused on three areas: information processing, emotional support, and visualization.}

\textcolor{black}{For information processing, AI tools assist with retrieving background information~\cite{pavel2015sceneskim}, ensuring narrative consistency~\cite{sanghrajka2017lisa, kapadia2015computer}, and organizing content~\cite{valls2016error, mateas2003experiment}. Tools like Final Draft~\cite{finaldraft} integrate AI for scene management and formatting, streamlining the screenwriting process. For emotional support, earlier studies have explored emotion analysis in narratives~\cite{stapleton2003interactive}. Goyal et al.'s AESOP system analyzes emotions through plot units, helping screenwriters understand emotional trajectories~\cite{goyal2010toward}. However, it relies on basic emotion tags (e.g., positive, negative, neutral), failing to capture complex emotional dynamics. Su et al.'s work on simulating basic and mixed emotions~\cite{su2007personality} improved character expression but lacked deeper insights into emotions tied to plans and reasoning, limiting its integration into screenwriting workflows. For visualization, previous research is divided into data visualization and visual representation. In data visualization, studies have summarized storylines~\cite{tapaswi2014storygraphs} and managed character arcs~\cite{kim2017visualizing}, providing a foundation for data-driven AI systems. In visual representation, AI tools use NLP to create 2D and 3D visualizations of screenplay content~\cite{10.1145/3172944.3172972, kim2017visualizing}, including character interactions~\cite{won2014generating} and scenes~\cite{hanser2009scenemaker}. However, these tools lack real-time feedback, personalized representation styles, and rely heavily on the quality of the screenplay content. Beyond these primary research areas, a recent work closely related to ours is the Dramatron system~\cite{10.1145/3544548.3581225}, which explores the integration of LLMs into screenwriting. However, it focuses solely on LLMs, neglecting the impact of AI-generated images and videos.}

\textcolor{black}{Additionally, previous empirical research has shown that AI can generate screenplay elements comparable to those created by human writers~\cite{ccelik4ai}, leveraging tools like ChatGPT~\cite{chatgpt, luchen2023chatgpt} and DeepStory~\cite{deepstory} to enhance efficiency, reduce costs~\cite{naji2024employing}, and inspire creativity~\cite{brako2023robots}. Despite these advancements, human creativity remains irreplaceable~\cite{naji2024employing, song2022analysis}. Meanwhile, ethical concerns, including biases~\cite{chow2020ghost} and copyright issues~\cite{kavitha2023copyright}, have sparked significant debate. However, these studies have not investigated how AI is integrated into specific stages of screenwriting and how screenwriters' attitudes concretely influence their practices.}


\textcolor{black}{To address these gaps, our study aims to examine screenwriters' current practices and attitudes toward AI integration at various stages within the workflow. Additionally, we investigate screenwriters’ expectations for future AI tools, considering both the potential of emerging AI technologies and possibilities beyond current advancements.} 
Ultimately, we provide suggestions for designing tailored human-AI co-creation tools that meet screenwriters' needs.

\begin{table*}
\centering
\footnotesize
\caption{Demographic Information of Participants. From left to right, each column presents the participant number, age, gender, background, years (Y) of screenwriting experience, training methods received, prior use of AI in screenwriting (yes/no), and their self-reported proficiency in using AI for screenwriting (5 = very proficient, 1 = not proficient at all). The specific training methods are represented by the following abbreviations in the table: institution courses (IC), classic scripts (CS), online videos (OV), instructor books (IB), and other (O).}
\Description{Description for Table 1:
The table displays demographic information of participants. It contains information organized by participant number (No.), age, gender, background (e.g., student, professional, enthusiast), years of screenwriting experience (Y), training methods received, prior use of AI in screenwriting (yes or no), and self-reported proficiency in using AI for screenwriting rated on a scale of 1 to 5 (5 being very proficient and 1 being not proficient at all).

Training methods are abbreviated as follows:
- IC: institution courses
- CS: classic scripts
- OV: online videos
- IB: instructor books
- O: other

The table lists 23 participants (P1–P18 and N1–N5), each row detailing:
1. Age range from 22 to 32.
2. Gender distribution includes male and female participants.
3. Backgrounds vary between students, professionals, and enthusiasts.
4. Years of screenwriting experience range from 0.5 years to 8 years.
5. Participants report training through a combination of methods like IC, CS, OV, IB, or O.
6. Prior use of AI in screenwriting shows some participants have used AI (marked "Yes") and others have not (marked "No").
7. Self-reported AI proficiency scores range from 1 (not proficient) to 5 (very proficient).} 

\label{tab:participant_data}
\begin{tabular}{|p{0.8cm}|p{0.8cm}|p{1.1cm}|p{1.7cm}|p{2.2cm}|p{2cm}|p{1.1cm}|p{2.5cm}|}

\hline
\textbf{No.} & \textbf{Age} & \textbf{Gender} & \textbf{Background}   & \textbf{Experience(Y)} & \textbf{Training}       & \textbf{Use AI} & \textbf{Proficiency of AI} \\ \hline
\textcolor{black}{P1}  & 24  & Female & Student      & 1             & IB, CS         & Yes    & 4                 \\ \hline
\textcolor{black}{P2}  & 26  & Male   & Professional & 7             & IC, CS         & Yes    & 4                 \\ \hline
\textcolor{black}{P3}  & 23  & Female & Enthusiast   & 0.5           & IC, CS, O      & Yes    & 3                 \\ \hline
\textcolor{black}{P4}  & 25  & Male   & Professional & 7             & IC, CS, OV     & Yes    & 4                 \\ \hline
\textcolor{black}{P5}  & 23  & Female & Student      & 5             & IC, IB, CS     & Yes    & 3                 \\ \hline
\textcolor{black}{P6}  & 24  & Female & Student      & 2             & IC, IB, CS     & Yes    & 3                 \\ \hline
\textcolor{black}{P7}  & 25  & Male   & Professional & 8             & IC, IB, CS     & Yes    & 4                 \\ \hline
\textcolor{black}{P8}  & 26  & Male   & Student      & 2             & IC, CS, OV, IB & Yes    & 5                 \\ \hline
\textcolor{black}{P9}  & 27  & Female & Enthusiast   & 1             & IC, CS, IB     & Yes    & 4                 \\ \hline
\textcolor{black}{P10} & 23  & Female & Student      & 5             & IC, CS, IB     & Yes    & 2                 \\ \hline
\textcolor{black}{P11} & 27  & Female & Professional & 6             & IC, IB, CS     & Yes    & 2                 \\ \hline
\textcolor{black}{P12} & 24  & Female & Enthusiast   & 0.5           & IC, CS, IB     & Yes    & 4                 \\ \hline
\textcolor{black}{P13} & 25  & Male   & Enthusiast   & 6             & IC, CS, IB     & Yes    & 4                 \\ \hline
\textcolor{black}{P14} & 27  & Male   & Professional & 2             & IC, CS, OV, IB & Yes    & 4                 \\ \hline
\textcolor{black}{P15} & 32  & Male   & Professional & 8             & IC, CS, OV, IB & Yes    & 5                 \\ \hline
\textcolor{black}{P16} & 23  & Female & Enthusiast   & 4             & IC, CS, IB     & Yes    & 4                 \\ \hline
\textcolor{black}{P17} & 23  & Female & Professional & 4             & IC, CS, OV     & Yes    & 4                 \\ \hline
\textcolor{black}{P18} & 22  & Female & Student      & 3             & IC, CS, IB     & Yes    & 3                 \\ \hline
\textcolor{black}{N1}  & 23  & Male   & Student      & 3             & IC, CS         & No     & 1                 \\ \hline
\textcolor{black}{N2}  & 22  & Female & Enthusiast   & 1             & CS             & No     & 1                 \\ \hline
\textcolor{black}{N3}  & 25  & Male   & Enthusiast   & 2             & IC, IB, CS     & No     & 1                 \\ \hline
\textcolor{black}{N4}  & 29  & Female & Professional & 6             & IC, CS, OV, IB & No     & 1                 \\ \hline
\textcolor{black}{N5}  & 23  & Female & Enthusiast   & 4             & IC, CS, IB     & No     & 1                 \\ \hline
\end{tabular}
    \label{tab:participants}
\end{table*}

\section{Methodology}


\textcolor{black}{Our study utilized a qualitative approach, conducting semi-structured interviews with 23 participants with screenwriting backgrounds. The objective of these interviews was to gain understanding into the integration of AI tools in screenwriting, with a particular focus on participants' usage, attitudes toward AI, and future expectations.}

 
\subsection{Participants}

Participants with screenwriting backgrounds, totaling 23, were recruited for this study through snowball sampling~\cite{goodman1961snowball}. All interviews were independently conducted by the first author. All participant information is based on self-reports: the gender distribution was 9 males and 14 females, with participants aged between 22 and 32 years (average age: 25). Their screenwriting backgrounds primarily consisted of students, enthusiasts, and professionals, with all participants having received professional training in screenwriting methods. On average, they had approximately four years of screenwriting-related experience, which included either studies, work, or a combination of both (refer to Table ~\ref{tab:participant_data}). Their average self-reported proficiency level with AI in screenwriting was three out of five, suggesting a moderate level of familiarity. The majority of participants (78\%) had experience using AI in screenwriting and are referred to as P (e.g., P1). The remaining 22\% had not used AI tools in screenwriting and are referred to as N (e.g., N1). We did not use prior experience with AI as an inclusion criterion during recruitment for the following reasons. Participants without AI experience offered valuable insights into most research questions, including workflows, challenges, attitudes, and expectations toward AI, thereby enriching the findings. We retained responses from all 23 participants to achieve more comprehensive results.
\textcolor{black}{Each interview was conducted individually via an online meeting platform to ensure participant engagement and thorough data collection.}

\subsection{Apparatus and Materials}


The study utilized a laptop and an online meeting system's integrated recording device. To explore the screenwriting process, we designed 27 open-ended questions, divided into three sections. The full list of questions is provided in the supplementary materials.

\begin{itemize}
\item Basic information and daily screenwriting workflow (\textbf{RQ1}) (e.g., screenplay types and themes, tools, workflow stages, and challenges)

\item Experience with AI tools in the screenwriting process (\textbf{RQ1}, \textbf{RQ2}) (e.g., familiarity and experience with AI, AI integration into workflow)

\item Discussion of potential AI assistance (\textbf{RQ3}) (e.g., ideal AI features, ideal interaction methods, visualization, role-playing)
\end{itemize}

The interviews began by focusing on participants' typical workflows, then explored their specific use and experiences with AI tools, and finally, inquired about their attitudes toward potential AI tools. The potential AI features mentioned in the interviews are based on previous related studies, and we aimed to understand participants' current views and future expectations for these features.

\subsection{Procedure}


The study followed widely recognized ethical frameworks~\cite{wma_helsinki_declaration}, prioritizing the safety, rights, and dignity of all participants. Participants were fully informed about the study’s background, objectives, and purpose prior to providing informed consent and participating in data collection. \textcolor{black}{Their rights were clearly explained, including the voluntary nature of participation, the option to withdraw at any time without penalty, and the assurance of confidentiality.} Privacy protection measures were explicitly outlined, such as anonymizing all responses and securely storing the collected data. To enable informed decision-making, participants were given one week to review the detailed consent form and carefully consider their participation, ensuring that no undue pressure or coercion was applied.

After obtaining informed consent, anonymized background information was collected from participants (see Table ~\ref{tab:participant_data}). Subsequently, in-depth, one-on-one interviews were conducted with each participant using a semi-structured format guided by a pre-prepared outline of 27 open-ended questions. Each interview lasted approximately 80 to 100 minutes, allowing participants to express their perspectives in detail. \textcolor{black}{To ensure accurate and comprehensive data collection, interviews were recorded using the online meeting platform.} All recordings were securely stored in a restricted-access environment and used solely for analysis. Throughout the research process, strict confidentiality was maintained, adhering to ethical standards and safeguarding participants’ privacy.

\begin{figure*}
 \centering         
\includegraphics[width=1\textwidth]{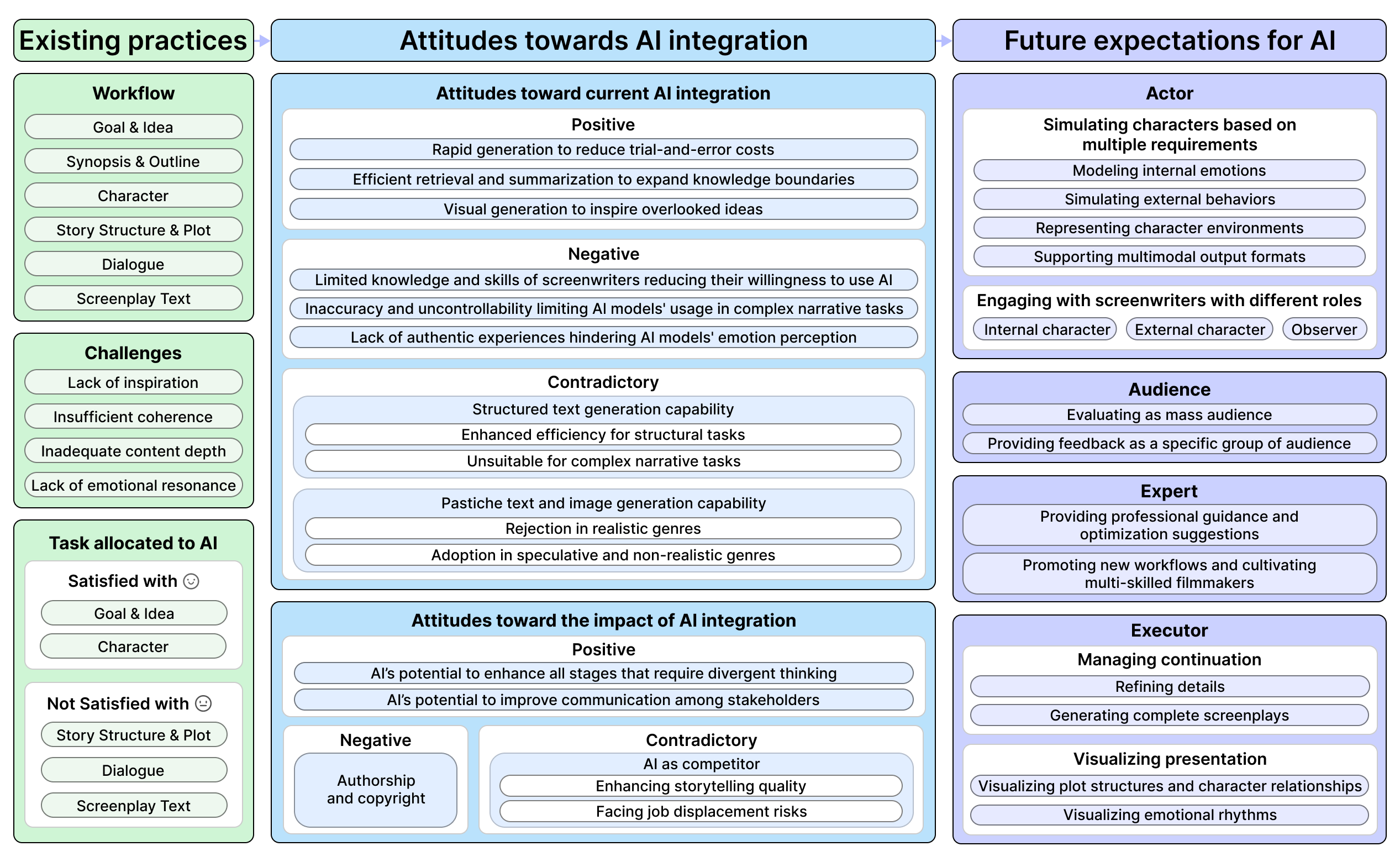} 
 \caption{\textcolor{black}{Overview of This Study’s Findings}. This figure summarizes the three themes aligned with the research questions: existing practices, \textcolor{black}{attitudes toward AI integration}, and future expectations for AI, including \textcolor{black}{nine} sub-themes and key findings.}
 \label{overview}
\Description{Description for Figure 1:
This Figureure summarizes the three themes aligned with the research questions: Existing practices, Attitudes towards AI integration, and Future expectations for AI, including nine sub-themes and all key findings.

Theme 1: Existing Practices
Sub-theme 1: Workflow: Screenwriters' creative process is divided into several steps, including:
  1. Goal and Idea
  2. Synopsis and Outline
  3. Character
  4. Story Structure and Plot
  5. Dialogue
  6. Screenplay Text
Sub-theme 2: Challenges faced by screenwriters during creation include:
  - Lack of Inspiration
  - Insufficient coherence
  - Inadequate content depth
  - Lack of emotional resonance
Sub-theme 3: Task Allocated to AI: Screenwriters' satisfaction with AI taking on certain tasks:
  - Satisfied with AI for Goal and Idea and Character stages.
  - Not satisfied with AI for Story Structure and Plot, Dialogue, and Screenplay Text stages.
Theme 2: Attitudes Towards AI
This section explores both positive and negative attitudes of screenwriters toward AI, divided as follows:
Sub-theme 4: Attitudes Toward Current AI Integration
1. Positive:
    - Rapid generation to reduce trial-and-error costs
- Efficient retrieval and summarization to expand knowledge boundaries
- Visual generation to inspire overlooked ideas
2. Negative
    - Limited knowledge and skills of screenwriters reducing their willingness to use AI
- Inaccuracy and uncontrollability limiting AI models' usage in complex narrative tasks
- Lack of authentic experiences hindering AI models' emotion perception 
3.Contradictory
-Structured text generation capability: Enhanced efficiency for structural tasks, and Unsuitable for complex narrative tasks
- Pastiche text and image generation capability: Rejection in realistic genres, and Adoption in speculative and non-realistic genres
Sub-theme 5: Attitudes Toward the Impact of AI Integration:
1. Positive:
    - AI’s potential to enhance all stages that require divergent thinking
- AI’s potential to improve communication among stakeholders
2.Negative:
- Authorship and copyright
3. Contradictory
- Al as competitor: Enhancing storytelling quality, and Facing job displacement risks
Theme 3: Future Expectations for AI
This section lists screenwriters' expectations for AI development in the future, divided into the following categories:
Sub-theme 6: Actor
1. Simulating characters based on multiple requirements:
  - Modeling internal emotions
  - Simulating external behaviors
  - Representing character environments
  - Supporting multimodal output formats
2. Engaging with screenwriters with different methods. Possible roles that screenwriters might take when interacting with AI:
  - Internal character
  - External character
  - Observer
Sub-theme 7: Audience
  - Evaluating as mass audience
  - Providing feedback as a specific group of audience
Sub-theme 8: Expert
  - Providing professional guidance and optimization suggestions
  - Promoting new workflows and cultivating multi-skilled filmmakers
Sub-theme 9: Executor
  - Managing continuation: Refining details, and Generating complete screenplays
- Visualizing presentation: Visualizing plot structures and character relationships, and Visualizing emotional rhythms}
 \end{figure*}

\subsection{Data Analysis}


Our analysis involved reviewing quotes extracted from approximately 1,992 minutes of transcribed audio from 23 participants. We collected 117 pages of interview notes and 558 pages of transcripts. Two authors conducted a qualitative analysis using an inductive approach to open coding, following the six stages of thematic analysis~\cite{braun2006using}: 1) Familiarizing ourselves with the data: We reviewed transcripts and recordings to ensure a full understanding of the content. 2) Generating initial codes: We independently coded sections based on the interview outline, including screenwriting workflow, AI tool feedback, and interaction methods. 3) Searching for themes: We compared initial codes and explored emerging themes. To adhere to the reflective thematic analysis method, we did not establish inter-rater reliability but instead acknowledged the influence of both authors on the analysis. 4) Reviewing themes: After multiple discussions, we refined the themes, aligning them with the data and identifying relevant examples. 5) Defining and naming themes: Themes were finalized, and their definitions were clarified. 6) Generating the report: The first author reviewed the final coding and discussed it with the other authors to produce the final report. This process led us to identify three themes with \textcolor{black}{nine} sub-themes, which will be presented in the following sections with detailed findings and examples as Fig. \ref{overview}:

\begin{itemize}
\item  \textbf{Section~\ref{sec:Practices}:} Existing practices: workflow, challenges, and task allocated to AI (\textbf{RQ1})

\item  \textbf{Section~\ref{sec:Attitudes}:} Attitudes: positive, negative, and contradictory cases (\textbf{RQ2})

\item  \textbf{Section~\ref{sec:Expectations}:} Future expectations: actor, audience, expert, and executor (\textbf{RQ3})

\end{itemize}

\section{\textcolor{black}{Findings 1: Existing Practices}}\label{sec:Practices}


Seventy-eight percent of our 23 participants reported having used AI in the screenwriting process, providing feedback on AI's application. Therefore, \textcolor{black}{Sections~\ref{sec:workflow} and~\ref{sec:challenges}} include perspectives from all 23 participants, while the \textcolor{black}{Section \ref{sec:Allocation}} reflects specific feedback from the 18 participants (78\%) who had used AI in screenwriting. The tools mentioned by participants fall into two categories: traditional screenwriting tools and AI tools. The AI tools they used include AI text generation tools (e.g., ChatGPT, WPS AI, Kimi, AI Dungeon), AI image generation tools (e.g., Midjourney, Runway), and AI sound generation tools (e.g., DeepMusic).

\subsection{Workflow}\label{sec:workflow}
\begin{figure*}
 \centering         
\includegraphics[width=1\textwidth]{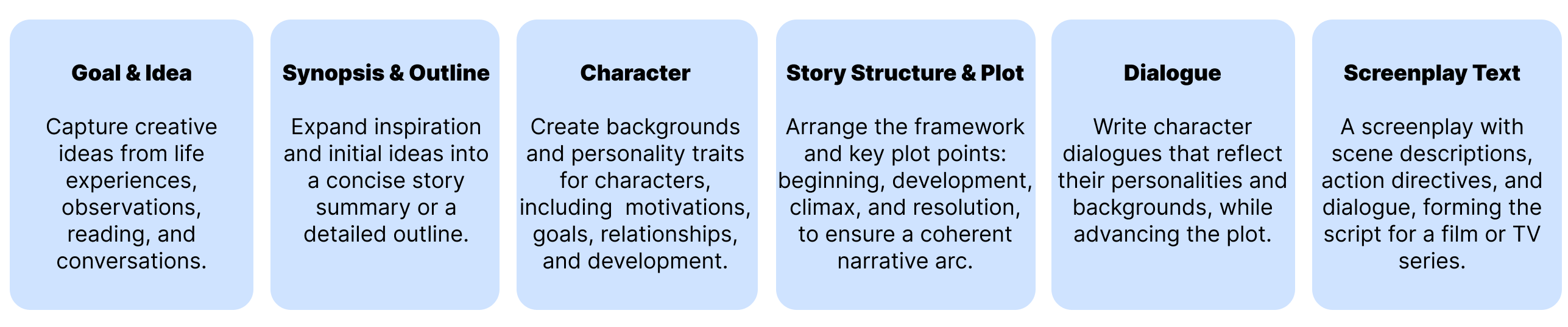} 
 \caption{A Common Screenwriting Workflow Summarized from 23 Participants. The six blue blocks represent common stages in the nonlinear workflow, without a specific order. The detailed workflows are provided in the supplementary material.}
 \label{workflow}
\Description{Description for Figure 2:
This diagram illustrates a common traditional screenwriting workflow, summarized from the practices of 23 participants. The process is represented by six blue blocks, each outlining a different stage in the nonlinear workflow. These stages can occur in any order, without a strict sequence. 

Here is a breakdown of each step in the process:

1. Goal & Idea: This stage focuses on capturing creative ideas and concepts through life experiences, observations, reading, and conversations.

2. Synopsis & Outline: The initial ideas and inspiration are expanded into a concise story summary or a more detailed outline.

3. Character: The development of characters involves creating backgrounds and personality traits, including their motivations, goals, relationships, and development arcs.

4. Story Structure & Plot: This step involves arranging the overall framework and key plot points, including the beginning, development, climax, and resolution, ensuring a coherent narrative arc.

5. Dialogue: Writing dialogues that reflect the characters' personalities and backgrounds, while also advancing the story.

6. Screenplay Text: A complete screenplay that includes scene descriptions, action directives, and dialogue, forming the full script for a film or TV series.}
 \end{figure*}
 

As mentioned in Section 2.4 regarding the complexity of the screenwriting workflow, we first summarized the participants' workflows. Through interviews with 23 screenwriters, we identified a common workflow encompassing the following key stages: goal \& idea, synopsis \& outline, character, story structure \& plot, dialogue, and screenplay text as Fig. \ref{workflow}. Despite individual differences, we found consensus on the core stages involved in screenwriting, highlighting shared practices across creative approaches. This commonality supports the workflow we have summarized. Our subsequent analysis will proceed within the context of this workflow.


\subsection{Challenges}\label{sec:challenges}
Before exploring how screenwriters use AI, we first analyzed the challenges they face in their current workflow, which can be categorized into four main aspects: lack of inspiration, insufficient coherence, inadequate content depth, and lack of emotional resonance. Among the participants, 15 out of 23 reported frequently experiencing a lack of inspiration, primarily due to a limited understanding of the real world and personal experience, which hinders the generation of innovative ideas in character development and plot creation. Additionally, 22 out of 23 participants mentioned issues with coherence in their screenplays. \textcolor{black}{The importance of aligning dialogue with context and character settings was emphasized by 13 participants, and 10 participants highlighted the need for logical coherence in plot structure.} Furthermore, seven out of 23 participants felt their screenplays often lacked depth, particularly in story themes and emotional dialogue. Another eight participants expressed difficulties in creating content that resonates with the audience, especially in dialogue and plot. In subsequent Sections~\ref{sec:Allocation}, we found that screenwriters have already begun exploring the use of AI to address some of these challenges.

\subsection{Task Allocated to AI} \label{sec:Allocation}

To understand screenwriters' use of AI across different workflow stages, we analyzed the usage and satisfaction across the six workflow stages mentioned above, based on responses from the \textcolor{black}{18} participants who had previously used AI in screenwriting, as shown in Table~\ref{tab:Task Allocation 23}. In addition, to further understand the tasks currently allocated to AI by screenwriters at different workflow stages, we categorized specific tasks by distinguishing between human-provided information and AI-generated content. Human-provided information refers to the input screenwriters contribute when interacting with AI, while AI-generated content refers to the output produced by AI based on the screenwriter's input, as outlined in Table~\ref{tab:Task Allocation}. \textcolor{black}{Note that the five participants who had not used AI in screenwriting (N) were not included in this section as they did not provide relevant data. We found that their reasons for not having tried AI in screenwriting were primarily due to perceiving AI as having a steep learning curve (N3, and N4) or assuming that AI-generated results would be unsatisfactory (N1, N2, N4, and N5).}

\newcolumntype{L}[1]{>{\raggedright\let\newline\\\arraybackslash\hspace{0pt}}m{#1}}
\newcolumntype{C}[1]{>{\centering\let\newline\\\arraybackslash\hspace{0pt}}m{#1}}
\newcolumntype{R}[1]{>{\raggedleft\let\newline\\\arraybackslash\hspace{0pt}}m{#1}}


\begin{table*}
\caption{\textcolor{black}{Task Allocation with AI in the Screenwriting Workflow Among 18 Screenwriters Who Have Previously Used AI in Screenwriting.} Green represents participants who used AI at this stage and were satisfied with AI output, \textcolor{black}{denoted as ``S'' in the table}. Red represents participants who used AI at this stage but were dissatisfied with AI output, \textcolor{black}{denoted as “D” in the table}. Blank indicates that the participants have not yet used AI tools at this stage.}
\Description{This table represents task allocation with AI in the screenwriting workflow among 18 screenwriters who have previously used AI in screenwriting. It provides information about whether participants used AI at different stages of the workflow and their satisfaction with the AI-generated output.
Key information:
  Rows represent workflow stages:
    1. Goal \& Idea
    2. Synopsis \& Outline
    3. Character
    4. Story Structure \& Plot
    5. Dialogue
    6. Screenplay Text
  Columns represent participants (P1 to P18), with a "Total" column summarizing usage counts for each stage.
Color codes and notations:
  Green cells marked as "S" indicate participants who used AI at a specific stage and were satisfied with its output.
  Red cells marked as "D" indicate participants who used AI at a specific stage but were dissatisfied with its output.
  Blank cells indicate that participants have not yet used AI tools at that stage.
Summary of usage:
1. Goal \& Idea: AI was used by 8 participants, all satisfied.
2. Synopsis \& Outline: AI was used by 3 participants, with mixed satisfaction levels.
3. Character: AI was used by 5 participants, with mixed satisfaction levels.
4. Story Structure \& Plot: AI was used by 12 participants, with mixed satisfaction levels.
5. Dialogue: AI was used by 6 participants, all dissatisfied.
6. Screenplay Text: AI was used by 9 participants, with mixed satisfaction levels.
The table highlights the stages where AI tools are used most often (e.g., Story Structure \& Plot) and participant satisfaction.}
\label{tab:Task Allocation 23}

\vspace{0.1cm}

\scriptsize


\newcommand{\patterncellred}{\cellcolor[HTML]{EF949F}\makebox[0pt][c]{D}D\tikz[overlay] \fill[pattern=north west lines, pattern color=black] (0,0) rectangle (\linewidth,\baselineskip);}
\newcommand{\patterncellgreen}{\cellcolor[HTML]{ADD88D}\makebox[0pt][c]{S}S\tikz[overlay] \fill[pattern=horizontal lines, pattern color=black] (0,0) rectangle (\linewidth,\baselineskip);}

\centering
\begin{tabular}{|c|c|c|c|c|c|c|c|c|c|c|c|c|c|c|c|c|c|c|c|}

\hline
\multicolumn{1}{|c|}{\textbf{Stage/Participant}} & \multicolumn{1}{c|}{\textcolor{black}{\textbf{P1}}} & \textcolor{black}{\textbf{P2}}                       & \textcolor{black}{\textbf{P3}}                       & \textcolor{black}{\textbf{P4}}                       & \textcolor{black}{\textbf{P5}}                       & \textcolor{black}{\textbf{P6}}                       & \textcolor{black}{\textbf{P7}}                       & \textcolor{black}{\textbf{P8}}                       & \textcolor{black}{\textbf{P9}}                       & \textcolor{black}{\textbf{P10}}                      & \textcolor{black}{\textbf{P11}}                      & \textcolor{black}{\textbf{P12}}                      & \textcolor{black}{\textbf{P13}}                      & \textcolor{black}{\textbf{P14}}                      & \textcolor{black}{\textbf{P15}}                      & \textcolor{black}{\textbf{P16}}                      & \textcolor{black}{\textbf{P17}}                      & \textcolor{black}{\textbf{P18}}                      & \textbf{Total} 

\\ \hline
Goal \& Idea            & \cellcolor[HTML]{ADD88D}\makebox[0pt][c]{S} &                          & \cellcolor[HTML]{ADD88D}\makebox[0pt][c]{S} & \cellcolor[HTML]{ADD88D}\makebox[0pt][c]{S}                        &                          &                          & \cellcolor[HTML]{ADD88D}\makebox[0pt][c]{S}                        &                          &                          &                          & \cellcolor[HTML]{ADD88D}\makebox[0pt][c]{S} &                          &                          & \cellcolor[HTML]{ADD88D}\makebox[0pt][c]{S}{\color[HTML]{588E31} } &                          & \cellcolor[HTML]{ADD88D}\makebox[0pt][c]{S}                        & \cellcolor[HTML]{ADD88D}\makebox[0pt][c]{S} &                          & 8     \\ \hline
Synopsis \& Outline     &                          &                          &                          & \cellcolor[HTML]{EF949F}\makebox[0pt][c]{D}                        & \cellcolor[HTML]{EF949F}\makebox[0pt][c]{D} &                          &                                                 &                          &                          &                          &                          &                          &                          &                                                 &                          &                                                 & \cellcolor[HTML]{ADD88D}\makebox[0pt][c]{S} &                          & 3     \\ \hline
Character               &                          &                          &                          & \cellcolor[HTML]{ADD88D}\makebox[0pt][c]{S}{\color[HTML]{ADD88D} } &                          &                          &                                                 &                          & \cellcolor[HTML]{ADD88D}\makebox[0pt][c]{S} &                          &                          &                          &                          &                                                 & \cellcolor[HTML]{ADD88D}\makebox[0pt][c]{S} & \cellcolor[HTML]{EF949F}\makebox[0pt][c]{D}                        & \cellcolor[HTML]{EF949F}\makebox[0pt][c]{D} &                          & 5     \\ \hline
Story Structure \& Plot & \cellcolor[HTML]{EF949F}\makebox[0pt][c]{D} & \cellcolor[HTML]{EF949F}\makebox[0pt][c]{D} &                          & \cellcolor[HTML]{EF949F}\makebox[0pt][c]{D}                        &                          & \cellcolor[HTML]{EF949F}\makebox[0pt][c]{D} &                                                 & \cellcolor[HTML]{ADD88D}\makebox[0pt][c]{S} & \cellcolor[HTML]{EF949F}\makebox[0pt][c]{D} &                          & \cellcolor[HTML]{EF949F}\makebox[0pt][c]{D} & \cellcolor[HTML]{EF949F}\makebox[0pt][c]{D} & \cellcolor[HTML]{EF949F}\makebox[0pt][c]{D} &                                                 & \cellcolor[HTML]{ADD88D}\makebox[0pt][c]{S} & \cellcolor[HTML]{EF949F}\makebox[0pt][c]{D}{\color[HTML]{588E31} } & \cellcolor[HTML]{EF949F}\makebox[0pt][c]{D} &                          & 12    \\ \hline
Dialoge                 & \cellcolor[HTML]{EF949F}\makebox[0pt][c]{D} &                          &                          & \cellcolor[HTML]{EF949F}\makebox[0pt][c]{D}{\color[HTML]{588E31} } &                          &                          &                                                 & \cellcolor[HTML]{EF949F}\makebox[0pt][c]{D} & \cellcolor[HTML]{EF949F}\makebox[0pt][c]{D} &                          & \cellcolor[HTML]{EF949F}\makebox[0pt][c]{D} &                          &                          &                                                 & \cellcolor[HTML]{EF949F}\makebox[0pt][c]{D} &                                                 &                          &                          & 6     \\ \hline
Screenplay Text         &                          &                          &                          & \cellcolor[HTML]{EF949F}\makebox[0pt][c]{D}                        &                          &                          & \cellcolor[HTML]{EF949F}\makebox[0pt][c]{D}{\color[HTML]{92D050} } &                          &                          & \cellcolor[HTML]{EF949F}\makebox[0pt][c]{D} &                          & \cellcolor[HTML]{EF949F}\makebox[0pt][c]{D} & \cellcolor[HTML]{EF949F}\makebox[0pt][c]{D} & \cellcolor[HTML]{EF949F}\makebox[0pt][c]{D}                        & \cellcolor[HTML]{EF949F}\makebox[0pt][c]{D} & \cellcolor[HTML]{ADD88D}\makebox[0pt][c]{S}                        & {\color[HTML]{ADD88D} }  & \cellcolor[HTML]{EF949F}\makebox[0pt][c]{D} & 9     \\ \hline
\end{tabular}
\end{table*}

\subsubsection{Goal \& Idea Stage}
In this stage, screenwriters often encounter challenges related to a lack of inspiration (Section~\ref{sec:challenges}). Consequently, they sought AI assistance. According to our interview results, P1, P3, P4, P7, P11, P14, P16, and P17 stated that they used AI at this stage and felt that AI could meet their needs. The human-provided information at this stage is primarily to determine the subject matter, genre, and elements to be included in the screenplay. Screenwriters use AI to obtain inspiration, assist with information retrieval, determine the theme direction, establish world-building, generate concept images, create story outlines, and develop character biographies. The needs of screenwriters at this stage are often vague, but they hope that AI can help them explore different concepts and inspire their next steps.

\subsubsection{Synopsis \& Outline Stage}
Only \textcolor{black}{P17} mentioned successfully using AI at this stage, stating that the human-provided information during use was script elements and character design allowing AI to generate synopsis \& outlines. 
\textcolor{black}{P17} stated, ``\textit{I sometimes feed the client's requirements into the AI to see what feedback it gives. It can help me set the tone by triggering certain keywords, which I then rearrange to get closer to what I want.}'' However, \textcolor{black}{P4 and P5} expressed dissatisfaction with AI's effectiveness at this stage. \textcolor{black}{P4} noted, ``\textit{At the beginning, I tried using keywords to generate an outline, and I also tried using an outline to generate a script, but both attempts failed. It felt like AI couldn’t understand the relationship between the characters. The content produced was strange and disconnected.}'' These responses indicate that AI, due to its lack of understanding of human emotions and complex human relationships, does not have sufficient capability to generate synopses and outlines. This aligns with the limitations in AI's current abilities (Section~\ref{sec:capabilities}).


\subsubsection{Character Stage}
\textcolor{black}{P4, P9, and P15} mentioned that their tasks using AI primarily involved providing character-related needs, such as naming characters and developing settings and script elements, with AI generating names and biographies based on their input. As \textcolor{black}{P4} noted: ``\textit{For example, when I suddenly needed a name, I used AI to generate a few names for me, and I thought they were quite interesting and playful.}'' \textcolor{black}{P15} stated: ``\textit{I had a spaceship captain character. I knew roughly his age and that he was a villain. The AI quickly gave me a detailed profile, including his height, weight, tragic childhood, and his current evil goals. I could then make modifications based on this. It was much faster than trying to come up with everything from scratch, especially since gathering material for this kind of character is so complex.}'' Participants indicated that, with current inputs, AI meets basic expectations for character-related tasks like generating names and biographies. However, when it comes to understanding relationships between characters, AI falls short (\textcolor{black}{P16 and P17}).

\subsubsection{Story Structure \& Plot Stage}
In this stage, screenwriters often meet challenges in maintaining the coherence of story pieces and having an in-depth structure (Section~\ref{sec:challenges}). As a result, they would like to seek AI's assistance. Compared to other stages, more screenwriters preferred to work with AI in this stage. 
According to our interview results, 12 participants reported using AI, but only two of them stated that AI could produce satisfactory results (\textcolor{black}{P8 and P15}). 
This stage had the most AI users and involved various human-provided informations, including providing genre, basic plots, storyboards, character settings, story outlines, scene settings, and reference images. 
AI tasks included expanding plot details, summarizing, organizing, continuing the plot, generating multiple plot possibilities, and generating multiple scene possibilities. Despite the specific and varied needs of screenwriters at this stage, \textcolor{black}{P1, P2, P4, P6, P9, P11, P12, P13, P16, and P17} expressed dissatisfaction with AI's performance. 
\textcolor{black}{P13} noted that while AI-generated structures were technically sound, the content was rigid. 
\textcolor{black}{P16} felt that while AI-generated images helped expand plot details, AI-generated text was often dull and logically incoherent. 
\textcolor{black}{P2, P6, and P17} criticized AI’s limited understanding of film and television resources, with \textcolor{black}{P17} stating: ``\textit{The content generated by the AI is unusable for me.}'' This is a process that participants frequently feel inefficient and time-consuming. These collective insights underscore the challenges screenwriters face when integrating AI into the story structure creation process, pointing to the need for significant improvements in AI's ability to understand and generate coherent, in-depth creative structure (detailed in Section~\ref{sec:capabilities}).

\subsubsection{Dialogue Stage}
At this stage, screenwriters often face challenges related to insufficient content depth and emotional resonance (Section~\ref{sec:challenges}), prompting them to explore the potential of AI assistance. However, P1, P4, P8, P9, P11, and P15 attempted to use AI for dialogue generation, but none were fully satisfied with its assistance. Their inputs include genre, character design, basic dialogue, and detailed dialogue scenarios. The primary concern among participants was that AI struggled with handling more complex scenarios, particularly in maintaining the nuanced interactions required for effective dialogue writing. For example, \textcolor{black}{P8} attempted to use ChatGPT to write a dialogue between two animals for a children’s story but found that AI couldn’t handle highly complex scenarios. \textcolor{black}{P11} mentioned, ``\textit{AI generated a short story with a Wong Kar-Wai style, including dialogue and a literary feel, but it was only useful for brief inspiration. It wasn’t suitable for actual screenwriting work.}'' These insights reveal a common sentiment among participants that, while AI can provide initial creative sparks, its current capabilities are inadequate for developing fully-realized dialogue in screenwriting, primarily due to its limited understanding of human emotions and complex interpersonal relationships. This observation aligns with the disadvantages of AI's current abilities as highlighted by participants (Section~\ref{sec:capabilities}).

\subsubsection{Screenplay Text Stage}
In this stage, screenwriters are often concerned about the inadequate content depth of the screenplay text (Section~\ref{sec:challenges}), leading 9 participants to experiment with using AI to assist in the entire screenplay creation process. However, only P16 felt that AI could optimize screenplay content, with \textcolor{black}{P16} noting that AI was effective in transforming plain sentences into narration in a specific style. The remaining participants (\textcolor{black}{P4, P7, P10, P12, P13, P14, P15, and P18}) were generally dissatisfied with AI’s performance. \textcolor{black}{P10} described AI-generated scripts as \textit{``completely useless and misleading.''} \textcolor{black}{P12} pointed out that AI only completed ``\textit{10\% to 20\% of the work, with the rest needing to be done manually.}'' \textcolor{black}{P18} also criticized AI’s output as ``\textit{too crude and not specific enough.}'' Based on the perspectives of our participants, the majority expressed that due to AI's current limitations in contextual understanding and logical coherence (Section~\ref{sec:capabilities}), it is generally unable to generate satisfactory screenplay text in most situations.

\subsubsection{Summary}
\textcolor{black}{In our study, screenwriters primarily used AI tools in four key stages: story structure \& plot development, screenplay text, goal \& idea generation, and dialogue, particularly for repetitive tasks like plot revisions (refer to Table~\ref{tab:Task Allocation 23} and Table~\ref{tab:Task Allocation}). However, they reported less AI use in the synopsis \& outline stages, preferring to rely on personal efforts. Regarding satisfaction, participants acknowledged AI’s ability to inspire ideas and meet basic needs in the goal \& idea generation and character development stages. However, fewer participants were satisfied with AI’s performance in the story structure \& plot development, screenplay text, and dialogue stages, due to its limitations (see Section~\ref{sec:capabilities}). Future research should consider enhancing AI capabilities in these areas.}

\begin{table*}
\caption{Task Allocated to AI in the Screenwriting Workflow}
\Description{Description for Table 3:
This Table provides a detailed breakdown of tasks allocated to AI in the screenwriting workflow, dividing responsibilities between human-provided information and AI-generated content across different workflow stages. The workflow stages are as follows:
1. Goal \& Idea
 Human-provided Information: Includes Genre (e.g., P3, P16) and Elements (e.g., P1, P3, P4, P7, P11, P14, P17).
 AI-generated Content: Provides Inspiration (P7, P14, P16), Assisted retrieval of information (P1, P4), Topic direction (P3), World-building (P17), Concept images (P11), Synopsis (P17), and Character biography (P17).
2. Synopsis \& Outline
 Human-provided Information: Includes Elements (P17) , Character design (P4), and Brief outline (P5).
 AI-generated Content: Generates Synopsis (P17) and Full outline (P4, P5).
3. Character
 Human-provided Information: Involves Naming requirements (P4), Character design (P15), and Elements (P9, P16, P17).
 AI-generated Content: Creates Character names (P4), Character biographies (P9, P15, P17), and Character relationship (P16, P17).
4. Story Structure \& Plot
 Human-provided Information: Includes Genre (P1), Basic plot (P2, P6, P8, P9, P15), Character design (P4, P6, P16, P17), Synopsis (P1, P12, P13, P16, P17), Scene design (P16), Scene reference image (P16), and Element (P16).
 AI-generated Content: Expands plot details (P1, P4, P9, P11, P12, P13), Plot continuation (P9, P16), Multiple plot possibilities (P1, P2, P6, P8, P9, P15, P17), and Multiple scene possibilities (P16).
5. Dialogue
 Human-provided Information: Covers Genre (P11), Character design (P4, P8, P9), Dialogue scenarios (P1, P4), Synopsis (P1), and No details were mentioned by P15.
 AI-generated Content: Includes Basic dialogue (P4, P8, P9), Stylized dialogue (P11), Continuation writing dialogue (P1, P9), and Generating complete dialogue (P1, P15).
6. Screenplay Text
 Human-provided Information: Includes Genre (P7, P10, P16, P18), Element ((P10, P16, P18), Screenplay text (P12, P16), Synopsis (P4, P13, P14), Character design (P18), and No details were mentioned by P15.
 AI-generated Content: Translation (P16), Narration (P16), Screenplay text (P4, P7, P10, P13, P14, P15, P18), and Format (P12).
This table highlights how human input guides AI outputs at each stage and shows specific participants associated with the contributions and outputs.}
\label{tab:Task Allocation}
\footnotesize
\begin{tabular}{|l|l|l|}
\hline
\textbf{Workflow Stage }                          & \textbf{Human-provided Information}                                                                                                                                                                                                                                                & \textbf{AI-generated Content}                                                                                                                                                                                                                                                   \\ \hline
\multirow{8}{*}{Goal \& Idea}            & \multirow{8}{*}{\begin{tabular}[c]{@{}l@{}}1. Genre (P3, P16)\\ 2. Element (P1, P3, P4, P7, P11, P14, P17)\end{tabular}}                                                                                                                                                  & \multirow{8}{*}{\begin{tabular}[c]{@{}l@{}}1. Inspiration (P7, P14, P16)\\ 2. Assisted retrieval of information (P1, P4)\\ 3. Topic direction (P3)\\ 4. World building (P17)\\ 5. Concept image (P11)\\ 6. Synopsis (P17)\\ 7. Character biography (P17)\end{tabular}} \\
                                         &                                                                                                                                                                                                                                                                           &                                                                                                                                                                                                                                                                        \\
                                         &                                                                                                                                                                                                                                                                           &                                                                                                                                                                                                                                                                        \\
                                         &                                                                                                                                                                                                                                                                           &                                                                                                                                                                                                                                                                        \\
                                         &                                                                                                                                                                                                                                                                           &                                                                                                                                                                                                                                                                        \\
                                         &                                                                                                                                                                                                                                                                           &                                                                                                                                                                                                                                                                        \\
                                         &                                                                                                                                                                                                                                                                           &                                                                                                                                                                                                                                                                        \\
                                         &                                                                                                                                                                                                                                                                           &                                                                                                                                                                                                                                                                        \\ \hline
\multirow{4}{*}{Synopsis \& Outline}     & \multirow{4}{*}{\begin{tabular}[c]{@{}l@{}}1. Element (P17)\\ 2. Character design(P4)\\ 3. Brief outline (P5)\end{tabular}}                                                                                                                                               & \multirow{4}{*}{\begin{tabular}[c]{@{}l@{}}1.Synopsis (P17)\\ 2.Full outline(P4, P5)\end{tabular}}                                                                                                                                                                     \\
                                         &                                                                                                                                                                                                                                                                           &                                                                                                                                                                                                                                                                        \\
                                         &                                                                                                                                                                                                                                                                           &                                                                                                                                                                                                                                                                        \\
                                         &                                                                                                                                                                                                                                                                           &                                                                                                                                                                                                                                                                        \\ \hline
\multirow{4}{*}{Character}               & \multirow{4}{*}{\begin{tabular}[c]{@{}l@{}}1. Naming requirements (P4)\\ 2. Character design (P15)\\ 3. Element (P9, P16, P17)\end{tabular}}                                                                                                                              & \multirow{4}{*}{\begin{tabular}[c]{@{}l@{}}1. Character name (P4)\\ 2. Character biography (P9, P15, P17)\\ 3. Character relationship (P16, P17)\end{tabular}}                                                                                                         \\
                                         &                                                                                                                                                                                                                                                                           &                                                                                                                                                                                                                                                                        \\
                                         &                                                                                                                                                                                                                                                                           &                                                                                                                                                                                                                                                                        \\
                                         &                                                                                                                                                                                                                                                                           &                                                                                                                                                                                                                                                                        \\ \hline
\multirow{8}{*}{Story Structure \& Plot} & \multirow{8}{*}{\begin{tabular}[c]{@{}l@{}}1. Genre (P11)\\ 2. Basic plot (P2, P6, P8, P9, P15)\\ 3. Character design (P4, P6, P16, P17)\\ 4. Synopsis (P1, P12, P13, P16, P17)\\ 5. Scene design (P16)\\ 6. Scene reference image (P16)\\ 7. Element (P16)\end{tabular}} & \multirow{8}{*}{\begin{tabular}[c]{@{}l@{}}1. Expand plot details (P1, P4, P9, P11, P12, P13)\\ 2. Plot continuation (P9, P16)\\ 3. Multiple plot possibilities (P1, P2, P6, P8, P9, P15, P17)\\ 4. Multiple scene possibilities (P16)\end{tabular}}                   \\
                                         &                                                                                                                                                                                                                                                                           &                                                                                                                                                                                                                                                                        \\
                                         &                                                                                                                                                                                                                                                                           &                                                                                                                                                                                                                                                                        \\
                                         &                                                                                                                                                                                                                                                                           &                                                                                                                                                                                                                                                                        \\
                                         &                                                                                                                                                                                                                                                                           &                                                                                                                                                                                                                                                                        \\
                                         &                                                                                                                                                                                                                                                                           &                                                                                                                                                                                                                                                                        \\
                                         &                                                                                                                                                                                                                                                                           &                                                                                                                                                                                                                                                                        \\
                                         &                                                                                                                                                                                                                                                                           &                                                                                                                                                                                                                                                                        \\ \hline
\multirow{6}{*}{Dialogue}                & \multirow{6}{*}{\begin{tabular}[c]{@{}l@{}}1. Genre (P11)\\ 2. Character design (P4, P8, P9)\\ 3. Dialogue scenarios (P1, P4)\\ 4. Synopsis (P1)\\ (No details were mentioned by P15)\end{tabular}}                                                                       & \multirow{6}{*}{\begin{tabular}[c]{@{}l@{}}1. Basic dialogue (P4, P8, P9)\\ 2. Stylized dialogue (P11)\\ 3. Continue writing dialogue (P1, P9)\\ 4. Generate complete dialogue (P1, P15)\end{tabular}}                                                                 \\
                                         &                                                                                                                                                                                                                                                                           &                                                                                                                                                                                                                                                                        \\
                                         &                                                                                                                                                                                                                                                                           &                                                                                                                                                                                                                                                                        \\
                                         &                                                                                                                                                                                                                                                                           &                                                                                                                                                                                                                                                                        \\
                                         &                                                                                                                                                                                                                                                                           &                                                                                                                                                                                                                                                                        \\
                                         &                                                                                                                                                                                                                                                                           &                                                                                                                                                                                                                                                                        \\ \hline
\multirow{7}{*}{Screenplay Text}         & \multirow{7}{*}{\begin{tabular}[c]{@{}l@{}}1. Genre (P7, P10, P16, P18)\\ 2. Element (P10, P16, P18)\\ 3. Screenplay text (P12, P16)\\ 4. Synopsis (P4, P13, P14)\\ 5. Character design (P18)\\ (No details were mentioned by P15)\end{tabular}}                          & \multirow{7}{*}{\begin{tabular}[c]{@{}l@{}}1. Translation (P16)\\ 2. Narration (P16)\\ 3. Screenplay text (P4, P7, P10, P13, P14, P15, P18)\\ 4. Format (P12)\end{tabular}}                                                                                            \\
                                         &                                                                                                                                                                                                                                                                           &                                                                                                                                                                                                                                                                        \\
                                         &                                                                                                                                                                                                                                                                           &                                                                                                                                                                                                                                                                        \\
                                         &                                                                                                                                                                                                                                                                           &                                                                                                                                                                                                                                                                        \\
                                         &                                                                                                                                                                                                                                                                           &                                                                                                                                                                                                                                                                        \\
                                         &                                                                                                                                                                                                                                                                           &                                                                                                                                                                                                                                                                        \\
                                         &                                                                                                                                                                                                                                                                           &                                                                                                                                                                                                                                                                        \\ \hline
\end{tabular}
\end{table*}

\section{\textcolor{black}{Findings 2: Attitudes Toward AI Integration}}\label{sec:Attitudes}



\textcolor{black}{Understanding participants' attitudes toward AI integration in screenwriting reveals its advantages and limitations across two dimensions: current AI integration, reflecting the perspectives of participants with prior AI experience, and broader impact, incorporating insights from all participants to provide a comprehensive view of AI's potential.}



\subsection{\textcolor{black}{Attitudes Toward Current AI Integration}}

\textcolor{black}{Participants with prior experience using AI in screenwriting, totaling 18, shared their perspectives on current AI capabilities.} These perspectives were categorized into three aspects: positive, negative, and contradictory. 
Additionally, the workflow stages discussed in this section align with those in Section~\ref{sec:Allocation} and  Tables~\ref{tab:Task Allocation 23} and~\ref{tab:Task Allocation}

\subsubsection{\textcolor{black}{Positive}} \label{sec:Current Positive}

\textcolor{black}{Many participants with prior experience using AI shared positive perspectives, which influenced their integration of AI into screenwriting workflows.}

\textcolor{black}{\textbf{Rapid Generation to Reduce Trial-and-Error Costs.}}
The value of AI’s ability to rapidly generate diverse ideas and narrative directions, particularly during the goal \& idea stage, was emphasized by 11 participants. P16 noted that AI’s brainstorming capabilities effectively addressed creative blocks, enabling screenwriters to explore new possibilities and refine initial concepts more efficiently. Similarly, participants (P8, and P15) highlighted AI’s utility in suggesting unconventional and alternative plotlines during the story structure \& plot stage, assisting in experimentation with innovative story forms and fresh perspectives.

\textcolor{black}{\textbf{Efficient Retrieval and Summarization to Expand Knowledge Boundaries.}}
AI's capability to retrieve and summarize diverse, unstructured information, especially during the goal \& idea generation and character development stages, was highlighted by 11 participants. P17 noted that AI tools efficiently reviewed extensive materials and extracted key insights from real-world research, providing valuable background information and allowing screenwriters to focus on refining creative details. P13 emphasized AI's potential for personalized knowledge expansion, stating, ``\textit{AI-generated visuals provide access to spaces difficult to explore in real life, like Mars or the universe... It expands our thinking and helps us beyond our knowledge}.'' P13 also noted that AI delivers more accurate and relevant content than traditional search engines when tailored to the screenplay's context, enhancing its utility.

\textcolor{black}{\textbf{Visual Generation to Inspire Overlooked Ideas.}}
P11, P14, and P16 also identified AI's ability to create visual imagery as a source of unexpected inspiration during the goal \& idea stage. P11 explained, ``\textit{AI generates concept art from just a few prompts, often revealing overlooked elements that inspire through color and composition},'' and emphasized the importance of focusing on elements beyond the initial prompt, which expands imaginative possibilities. P14 added, ``\textit{AI-generated visuals, such as character expressions or clothing, offer perspectives I would never consider},'' highlighting how visualizations allow screenwriters to preview potential visual effects and uncover overlooked creative opportunities.

\subsubsection{\textcolor{black}{Negative}} \label{sec:capabilities}

\textcolor{black}{All participants with prior AI experience expressed concerns about its capabilities, which impacted their use of AI in screenwriting workflows.}

\textcolor{black}{\textbf{Limited Knowledge and Skills of Screenwriters Reducing Their Willingness to Use AI.}}
\textcolor{black}{P1, P4, P8, P17, and P18 identified high barriers to effectively using AI due to their limited training, aligning with challenges noted by N3 and N4 in Section~\ref{sec:Allocation}. P17 observed, ``\textit{We lack a systematic method to train AI. It may have the capability we need, but we fail to articulate our requirements clearly. It is less about training AI and more about training ourselves}.'' These challenges reflect participants’ limited knowledge and skills in leveraging AI technologies effectively.}

\textcolor{black}{\textbf{Inaccuracy and Uncontrollability Limiting AI Models' Usage in Complex Narrative Tasks.}}
\textcolor{black}{P1, P3, P8, P9, P10, P11, P16, P17, and P18 cited AI’s inaccuracy and lack of control as major challenges, echoing concerns raised by N1, N2, and N5 in Section~\ref{sec:Allocation}. P8, P10, and P18 avoided using AI for critical tasks like complex story development, fearing misinformation. P10 noted, ``\textit{When analyzing Kubrick’s A Clockwork Orange, AI misrepresented his style and falsely attributed films by other directors to him}.'' These concerns highlight mistrust in AI’s reliability for tasks requiring accuracy and factual consistency.}


\textcolor{black}{\textbf{Lack of Authentic Experiences Hindering AI Models' Emotion Perception.}}
\textcolor{black}{AI lacks the ability to replicate authentic experiences, which are deeply tied to personal emotions and life events, as noted by 12 participants.} P2 explained, ``\textit{Creativity stems from my emotions—what makes me deeply pained or joyful.}'' P4, P7, and P8 avoided AI for tasks demanding deep emotional resonance, such as pivotal scenes. P7 observed, ``\textit{AI captures simple emotions, like a father’s love depicted as both smiling happily. But love is far more complex than that.}'' Participants highlighted that human creativity integrates nuanced, dynamic emotional flows across dialogue, behavior, and context, whereas AI relies on simplistic, discrete representations, limiting its application in screenwriting.

\subsubsection{Contradictory}

\textcolor{black}{``Contradictory'' refers to the varying attitudes expressed by participants toward certain AI capabilities, depending on usage contexts.} From our findings, feedback from 18 participants revealed mixed opinions on structured and pastiche generation, influencing the integration of AI into workflows.

\textbf{Structured Text Generation Capability.}
Participants argued that AI provides \textit{enhanced efficiency for structural tasks}. P2, P8, P12, P15, and P17 expressed positive views on structured text generation, highlighting its utility for predefined tasks such as screenplay outlines or foundational structures. P12 described that AI can assist us more with format-related tasks. P15 emphasized its value for beginners, as it reduces cognitive load and allows for a greater focus on creative aspects. 
On the other hand, participants also expressed that AI is \textit{unsuitable for complex narrative tasks.} P1, P12, P13, P16, and P18, who were critical of structured text generation, pointed out that it is overly structure-driven, resulting in repetitive frameworks and limited meaningful plot progression. They deemed it unsuitable for dialogue writing or intricate narrative development within the complete screenplay text.

\textbf{Pastiche Text and Image Generation Capability.}
Participants mentioned that AI has been \textit{adopted in speculative and non-realistic genres}. P9, P11, P12, P14, and P17 valued AI's imaginative potential for idea generation in science fiction and surrealism. P14 remarked, ``\textit{AI is good at piecing together different elements and creating connections and narrative forms that differ from human thinking},'' enabling the exploration of unconventional connections and narrative styles well-suited to these genres. On the contrary, participants highlighted their \textit{rejection of AI usage in realistic genres}. P4, P7, P8, P9, P12, P14, P16, and P17 avoided AI-generated pastiche content in logically coherent genres like documentaries. P4 and P16 criticized AI's poor contextual understanding, which caused illogical plot twists and unrelated character behaviors, disrupting the storyline.

\subsection{\textcolor{black}{Attitudes Toward the Impact of AI Integration}}  

\textcolor{black}{When discussing broader attitudes toward the impact of AI integration, all 23 participants shared their perspectives, which were also categorized into three aspects: positive, negative, and contradictory.}

\subsubsection{\textcolor{black}{Positive}} \label{sec:Positive Potential} 

\textcolor{black}{Participants anticipated increased human-AI collaboration across various stages of screenwriting workflows and the broader development of the film industry.}

\textcolor{black}{\textbf{AI’s Potential to Enhance All Stages that Require Divergent Thinking}.}
Most participants regarded AI as a creative collaborator. P8 noted its potential to generate diverse dialogues efficiently by understanding scenarios, particularly for dialogue and screenplay text stages. Even participants without AI experience expressed optimism, suggesting AI could assist in scene visualization (N1) and character interactions (N3) for dialogue creation. \textcolor{black}{AI was also perceived as enhancing goal and idea development, as well as story structure and plot creation, both driven by divergent thinking.}

\textcolor{black}{\textbf{AI’s Potential to Improve Communication among Stakeholders}.}
\textcolor{black}{P4, P7, P8, P13, P14, and P15 highlighted AI's potential to improve efficiency and reduce costs by facilitating better stakeholder communication. For instance, P13 suggested that AI-generated visual previews could be used to pitch screenplays or outlines to investors and directors. Such features were seen as tools to streamline workflows and foster collaboration as the industry increasingly adopts AI innovations.}

\subsubsection{\textcolor{black}{Negative}}  \label{sec:ethical concerns}  

\textcolor{black}{Participants expressed concerns about ethical issues surrounding AI, which influenced their attitudes toward its impact. Three participants specifically raised concerns about \textbf{authorship and copyright}. N2 questioned whether AI should be acknowledged as the true author, while P8 and P17 discussed copyright implications. P8 contended that screenwriters should retain copyright when AI serves merely as a framework provider. P17 suggested that iterative refinement with user input could produce distinct outcomes. However, P17 also noted that some colleagues equated such processes with plagiarism. These unresolved issues of ownership, originality, and collaboration contributed to participants' hesitancy to fully adopt AI tools.}

\begin{table*}

\caption{Summary of the Four Roles of AI in Screenwriters' Expectations}
\Description{Description for Table 4:
This Table provides a summary of the four roles of AI as expected by screenwriters. It consists of two columns: the role and its definition. Each role highlights how AI is perceived to assist screenwriters in different capacities:
1. Actor: AI is expected to enhance creative abilities by embodying characters and assisting screenwriters in expanding their knowledge boundaries.
2. Audience: AI aids screenwriters by evaluating the value of their creations from the perspective of diverse audience groups, thereby increasing acceptance and recognition of their work.
3. Expert: AI is seen as an authoritative guide, offering evaluations and suggestions while introducing new workflows based on professional expertise.
4. Executor: AI fulfills creative tasks according to specified demands, focusing on improving work efficiency.}
\label{tab:role}

\begin{tabular}{|p{1.5cm}|p{15cm}|}
     \hline
\textbf{Role}     & \textbf{Definition}                                                                                                                                          \\ \hline
Actor    & Enhancing creative abilities by embodying characters and assisting screenwriters in expanding knowledge boundaries.                           \\ \hline
Audience & Assisting screenwriters in evaluating the value of their creations from the perspective of diverse audience groups, increasing the acceptance and recognition of their work. \\ \hline
Expert   & Offering authoritative evaluations and suggestions, guiding new workflows through professional expertise.                                           \\ \hline
Executor & Fulfilling creative tasks according to specified demands, improving work efficiency.                                                                \\ \hline
\end{tabular}
\end{table*}

\subsubsection{\textcolor{black}{Contradictory}}  \label{sec:Contradictory AI as a competitor}
\textcolor{black}{Participants expressed conflicting views on \textbf{AI as a competitor.}
Those with positive opinions believed that AI has the potential to \textit{\textcolor{black}{enhance storytelling quality.}} \textcolor{black}{P4, P7, P13, P14, and P18 believed AI could stimulate competition and contribute to enhanced content quality, with P13 likening it to an ``\textit{electronic catfish}'' that encourages better storytelling. P18 highlighted that continuous advancements in AI techniques signify societal progress.} However, participants also thought there was a possibility of AI causing screenwriters to \textit{\textcolor{black}{face job displacement risks.}} \textcolor{black}{P1, P4, P6, P7, P13, and P18 highlighted concerns about job displacement due to AI. P7 specifically expressed fears about the potential erosion of professional skills, which could escalate into unemployment risks as AI continues to advance. P7 noted, ``\textit{If you delegate simple tasks to AI, you may lose the ability to handle complex tasks over time.}''}}

\section{\textcolor{black}{Findings 3: Future Expectations}}\label{sec:Expectations}

Participants expressed varied expectations for AI in screenwriting, and we categorized these future expectations into four distinct roles for AI: actor, audience, expert, and executor (refer to Table~ \ref{tab:role}). \textcolor{black}{To further understand how screenwriters' expectations of these roles influence their workflow stages, we also mapped each specific expectation to the stages where participants envisioned these expectations could be applied (refer to Table~ \ref{tab:expectations and stages})}.

\begin{table*}
\caption{\textcolor{black}{Summary of Screenwriters' Expectations for Four Roles and Their Potential Application to Workflow Stages. Blue cells indicate the stages where participants expected these roles to be applicable, denoted as ``E'' in the table. Blank cells represent stages where no application was mentioned. For simplicity, this table uses ``Idea'' to represent ``Goal \& Idea,'' ``Outline'' to represent ``Synopsis \& Outline,'' ``Plot'' to represent ``Story Structure \& Plot,'' and ``Screenplay'' to represent ``Screenplay Text.''}}
\scalebox{0.88}{
\Description{This Table summarizes screenwriters' expectations for four roles of AI (Actor, Audience, Expert, Executor) and their potential application to workflow stages. Blue cells indicate the stages where participants expected these roles to be applicable, denoted as ''E'' in the table. Blank cells represent stages where no application was mentioned:
  "Idea" for Goal \& Idea,
  "Outline" for Synopsis \& Outline,
  "Character" for Character,
  "Plot" for Story Structure \& Plot,
  "Dialogue" for Dialogue,
  "Screenplay" for Screenplay Text.
Role and Workflow Stage Mapping:
1. Actor:
Simulating characters based on multiple requirements:
Modeling internal emotions: Applicable to Idea, Character, Plot and Dialogue stages.
          Simulating external behaviors: Applicable to Idea, Character, Plot, Dialogue and Screenplay stages.
          Representing character environments: Applicable to Idea, Character, Plot, Dialogue and Screenplay stages.
          Supporting multimodal output formats: Applicable to Idea, Character, Plot, Dialogue and Screenplay stages.
Engaging with screenwriters with different methods:
          Internal character: Applicable to Idea, Character and Dialogue stages.
          External character: Applicable to Idea, Character, Plot and Dialogue stages.
          Observer: Applicable to Idea, Character, Plot, Dialogue and Screenplay stages.
2. Audience:
Evaluating as mass audience: Applicable to Idea, Character, Plot and Screenplay stages.
Providing feedback as a specific group of audience: Applicable to Idea stage.
3. Expert:
Providing professional guidance and optimization suggestions: Applicable to all stages.
Promoting new workflows and cultivating filmmakers: Applicable to Idea, Plot and Screenplay stages.
4. Executor:
Managing continuation: 
          Refining details: Applicable to all stages.
          Generating complete screenplays: Applicable to Screenplay stage.
Visualizing Presentation:
          Visualizing plot structures and character relationships: Applicable to Outline, Character, Plot and Screenplay stages.
          Visualizing emotional rhythms: Applicable to Character and Plot stage.
This table provides a comprehensive mapping of AI roles to various stages, showcasing how different stages of screenwriting workflow can benefit from AI involvement.}
\label{tab:expectations and stages}
\footnotesize
\centering

\begin{tabular}{|cll|c|c|c|c|c|c|}
\hline
\multicolumn{3}{|c|}{\textbf{Expectation/Stage}}                                                                                                                                                                  & \textbf{Idea}                                            & \textbf{Outline}                  & \textbf{Character}                                       & \textbf{Plot}                     & \textbf{Dialogue}                 & \textbf{Screenplay}               \\ \hline
\multicolumn{1}{|c|}{}                           & \multicolumn{1}{l|}{}                                                                       & Modeling internal emotions                              & \cellcolor[HTML]{CEDCFF}\makebox[0pt][c]{E}                        &                          & \cellcolor[HTML]{CEDCFF}\makebox[0pt][c]{E}                        & \cellcolor[HTML]{CEDCFF}\makebox[0pt][c]{E} & \cellcolor[HTML]{CEDCFF}\makebox[0pt][c]{E} &                          \\ \cline{3-3}
\multicolumn{1}{|c|}{}                           & \multicolumn{1}{l|}{}                                                                       & Simulating external behaviors                           & \cellcolor[HTML]{CEDCFF}\makebox[0pt][c]{E}                        &                          & \cellcolor[HTML]{CEDCFF}\makebox[0pt][c]{E}                        & \cellcolor[HTML]{CEDCFF}\makebox[0pt][c]{E} & \cellcolor[HTML]{CEDCFF}\makebox[0pt][c]{E} & \cellcolor[HTML]{CEDCFF}\makebox[0pt][c]{E} \\ \cline{3-3}
\multicolumn{1}{|c|}{}                           & \multicolumn{1}{l|}{}                                                                       & Representing character environments                     & \cellcolor[HTML]{CEDCFF}\makebox[0pt][c]{E}                        &                          & \cellcolor[HTML]{CEDCFF}\makebox[0pt][c]{E}                        & \cellcolor[HTML]{CEDCFF}\makebox[0pt][c]{E} & \cellcolor[HTML]{CEDCFF}\makebox[0pt][c]{E} & \cellcolor[HTML]{CEDCFF}\makebox[0pt][c]{E} \\ \cline{3-3}
\multicolumn{1}{|c|}{}                           & \multicolumn{1}{l|}{\multirow{-4}{*}{Simulating characters based on multiple requirements}} & Supporting multimodal output formats                    & \cellcolor[HTML]{CEDCFF}\makebox[0pt][c]{E}                        &                          & \cellcolor[HTML]{CEDCFF}\makebox[0pt][c]{E}                        & \cellcolor[HTML]{CEDCFF}\makebox[0pt][c]{E} & \cellcolor[HTML]{CEDCFF}\makebox[0pt][c]{E} & \cellcolor[HTML]{CEDCFF}\makebox[0pt][c]{E} \\ \cline{2-3}
\multicolumn{1}{|c|}{}                           & \multicolumn{1}{l|}{}                                                                       & Internal character                                      & \cellcolor[HTML]{CEDCFF}\makebox[0pt][c]{E}                        &                          & \cellcolor[HTML]{CEDCFF}\makebox[0pt][c]{E}                        &                          & \cellcolor[HTML]{CEDCFF}\makebox[0pt][c]{E} &                          \\ \cline{3-3}
\multicolumn{1}{|c|}{}                           & \multicolumn{1}{l|}{}                                                                       & External character                                      & \cellcolor[HTML]{CEDCFF}\makebox[0pt][c]{E}                        &                          & \cellcolor[HTML]{CEDCFF}\makebox[0pt][c]{E}                        & \cellcolor[HTML]{CEDCFF}\makebox[0pt][c]{E} & \cellcolor[HTML]{CEDCFF}\makebox[0pt][c]{E} &                          \\ \cline{3-3}
\multicolumn{1}{|c|}{\multirow{-7}{*}{Actor}}    & \multicolumn{1}{l|}{\multirow{-3}{*}{Engaging with screenwriters with different methods}}     & Observer                                                & \cellcolor[HTML]{CEDCFF}\makebox[0pt][c]{E}{\color[HTML]{ADD88D} } &                          & \cellcolor[HTML]{CEDCFF}\makebox[0pt][c]{E}                        & \cellcolor[HTML]{CEDCFF}\makebox[0pt][c]{E} & \cellcolor[HTML]{CEDCFF}\makebox[0pt][c]{E} & \cellcolor[HTML]{CEDCFF}\makebox[0pt][c]{E} \\ \hline
\multicolumn{1}{|c|}{}                           & \multicolumn{2}{l|}{Evaluating as mass audience}                                                                                                      & \cellcolor[HTML]{CEDCFF}\makebox[0pt][c]{E}                        &                          & \cellcolor[HTML]{CEDCFF}\makebox[0pt][c]{E}                        & \cellcolor[HTML]{CEDCFF}\makebox[0pt][c]{E} &                          & \cellcolor[HTML]{CEDCFF}\makebox[0pt][c]{E} \\ \cline{2-3}
\multicolumn{1}{|c|}{\multirow{-2}{*}{Audience}} & \multicolumn{2}{l|}{Providing feedback as a specific group of audience}                                                                               & \cellcolor[HTML]{CEDCFF}\makebox[0pt][c]{E}                        &                          &                                                 &                          &                          &                          \\ \hline
\multicolumn{1}{|c|}{}                           & \multicolumn{2}{l|}{Providing professional guidance and optimization suggestions}                                                                     & \cellcolor[HTML]{CEDCFF}\makebox[0pt][c]{E}                        & \cellcolor[HTML]{CEDCFF}\makebox[0pt][c]{E} & \cellcolor[HTML]{CEDCFF}\makebox[0pt][c]{E}                        & \cellcolor[HTML]{CEDCFF}\makebox[0pt][c]{E} & \cellcolor[HTML]{CEDCFF}\makebox[0pt][c]{E} & \cellcolor[HTML]{CEDCFF}\makebox[0pt][c]{E} \\ \cline{2-3}
\multicolumn{1}{|c|}{\multirow{-2}{*}{Expert}}   & \multicolumn{2}{l|}{Promoting new workflows and cultivating multi-skilled filmmakers}                                                                 & \cellcolor[HTML]{CEDCFF}\makebox[0pt][c]{E}                        &                          &                                                 & \cellcolor[HTML]{CEDCFF}\makebox[0pt][c]{E} &                          & \cellcolor[HTML]{CEDCFF}\makebox[0pt][c]{E} \\ \hline
\multicolumn{1}{|c|}{}                           & \multicolumn{1}{c|}{}                                                                       & Refining details                                        & \cellcolor[HTML]{CEDCFF}\makebox[0pt][c]{E}                        & \cellcolor[HTML]{CEDCFF}\makebox[0pt][c]{E} & \cellcolor[HTML]{CEDCFF}\makebox[0pt][c]{E}                        & \cellcolor[HTML]{CEDCFF}\makebox[0pt][c]{E} & \cellcolor[HTML]{CEDCFF}\makebox[0pt][c]{E} & \cellcolor[HTML]{CEDCFF}\makebox[0pt][c]{E} \\ \cline{3-3}
\multicolumn{1}{|c|}{}                           & \multicolumn{1}{c|}{\multirow{-2}{*}{Managing continuation}}                                & Generating complete screenplays                         &                                                 &                          &                                                 &                          &                          & \cellcolor[HTML]{CEDCFF}\makebox[0pt][c]{E} \\ \cline{2-3}
\multicolumn{1}{|c|}{}                           & \multicolumn{1}{c|}{}                                                                       & Visualizing plot structures and character relationships &                                                 & \cellcolor[HTML]{CEDCFF}\makebox[0pt][c]{E} & \cellcolor[HTML]{CEDCFF}\makebox[0pt][c]{E}                        & \cellcolor[HTML]{CEDCFF}\makebox[0pt][c]{E} &                          & \cellcolor[HTML]{CEDCFF}\makebox[0pt][c]{E} \\ \cline{3-3}
\multicolumn{1}{|c|}{\multirow{-4}{*}{Executor}} & \multicolumn{1}{c|}{\multirow{-2}{*}{Visualizing presentation}}                             & Visualizing emotional rhythms                           &                                                 &                          & \cellcolor[HTML]{CEDCFF}\makebox[0pt][c]{E}{\color[HTML]{ADD88D} } & \cellcolor[HTML]{CEDCFF}\makebox[0pt][c]{E} &                          &                          \\ \hline
\end{tabular}
}
\end{table*}

\subsection{Actor: Enhancing Creative Abilities by Embodying Characters and Assisting Screenwriters in Expanding Knowledge Boundaries}
All 23 participants expressed future expectations for AI-related in the role of an ``actor.'' Traditional screenwriting methods typically rely on the screenwriter's personal experience and relevant materials. However, due to the limitations of their knowledge base, participants sought a reliable source of inspiration to deepen their understanding of character identities and traits. To address these needs, we categorized the screenwriters' expectations of AI as an ``actor'' into two key areas: AI functions and screenwriter engagement. \textcolor{black}{Additionally, as participants perceived the ``actor'' as being related to and aligned with the screenplay's characters, this role was expected to provide more detailed character information, supporting the exploration of the screenplay's goals, along with the characters' motivations and emotions. Participants anticipated that the ``actor'' role would be primarily applied during the goal \& idea, character, and dialogue stages (as shown in Table~\ref{tab:expectations and stages}).}

\subsubsection{Simulating Characters Based on Multiple Requirements} \label{sec:Simulating}

Many participants argue that screenplay logic is inherently visual, and AI should use both text and visual representations when generating character actions and emotional responses. Excluding N2, the remaining 22 participants expect AI to generate detailed character images, including behaviors, actions, emotional changes, and complex relationships while interacting with screenwriters in real-time. This would help screenwriters intuitively understand characters' internal emotions and external behaviors, making the characters more vivid and multidimensional.

\textbf{Modeling Internal Emotions.} \textcolor{black}{P2, P4, P5, P11, P13, P14, P17, and N4 emphasized} that AI has the potential to stimulate creative inspiration by consistently portraying specific emotional performances. In particular, \textcolor{black}{P17} argued that AI's real-time display of facial expressions and psychological states could provide screenwriters with deeper insights into a character's inner world, enabling more precise adjustments to the screenplay's rhythm and enhancing the depth of the characters' internal monologues.


\textbf{Simulating External Behaviors.} P2, P4, P5, P6, P9, P10, P11, P12, P13, P14, P15, P16, and N4 expressed needs for the portrayal of character behaviors in the screenplay. \textcolor{black}{P12} suggested that AI could reference existing films of similar types to provide valuable scenes and emotional expression patterns for the screenplay creation. \textcolor{black}{P16} mentioned that by dialoguing with AI-embodied characters, screenwriters could explore how characters react in specific situations and incorporate these reactions into the screenplay. \textcolor{black}{P2 and P13} argue that AI can simulate characters' behaviors in various backgrounds, enabling screenwriters to understand and express the emotional complexity of characters more comprehensively without the limitations of geography and resources.

\textbf{Representing Character Environments.} \textcolor{black}{P2, P3, P5, P6, P8, P10, P11, P12, P13, P15, P16, N1, N3, and N4} mentioned that traditional screenwriting relies on imagination, whereas AI could help screenwriters visually present their ideas by generating scenes and character images, thereby enhancing the efficiency of both creation and communication during teamwork. Although current AI image generation tools can generate scenes to some extent, many participants expressed higher expectations for AI’s ability to generate the structure of the space, local environment, scene layout, and final imagery in which characters engage in dialogue.


\textbf{Supporting Multimodal Output Formats.} P2, P3, P4, P6, P10, P11, P12, P13, P14, P15, P16, N1, N3, and N5 strongly expected multimodal output, such as video generation, image generation, text generation, and speech generation. Among these participants, 10 favored output combining text with image and video generation. \textcolor{black}{P16} emphasized the advantages of integrating multimodal information, stating, ``\textit{It’s best to provide a variety of information, such as text, action descriptions, images, dynamic videos, etc., which can present more detailed scenes and help screenwriters better understand and create.}'' Meanwhile, \textcolor{black}{P15, N1, and N5} specifically mentioned the importance of combining text and speech output. \textcolor{black}{P15} noted, ``\textit{I think the story is already deep enough through text, but if understanding is difficult, we can supplement with sound. Even without visuals, you can still feel the emotion through sound.}'' Furthermore, \textcolor{black}{P2, P13, and P16} highlighted AI’s potential in offering immersive experiences and enhancing reality perception. \textcolor{black}{P2} stated, ``\textit{I prefer it to be 3D projection right next to me.}''

\subsubsection{Engaging with Screenwriters with Different Methods.} \label{sec:Engaging}
\begin{figure*}
 \centering         
\includegraphics[width=1\textwidth]{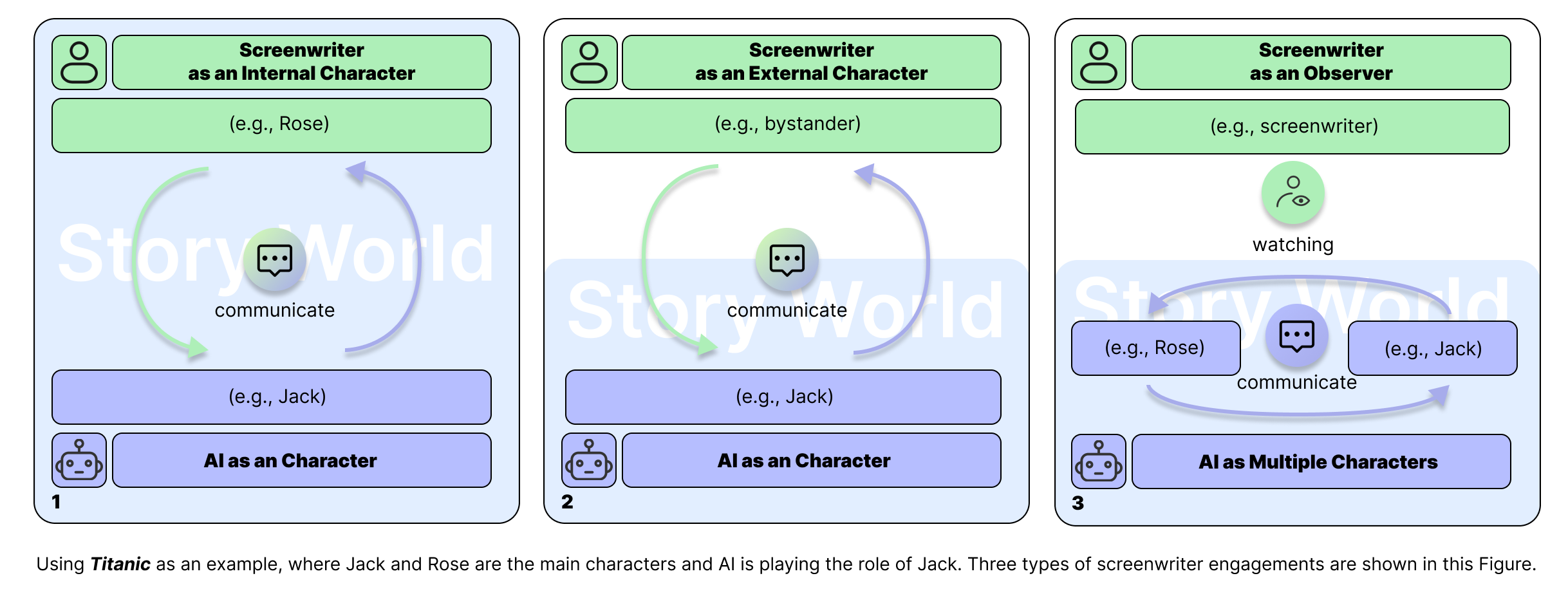} 
 \caption{The Three Categories of Screenwriter Engagement with AI Actors}
 \Description{Description for Figure 3:
This chart illustrates three different modes of screenwriter engagement with AI during the creative process: Internal Character, External Character, and Observer. Using the characters Jack and Rose from Titanic as examples, the chart explains how screenwriters interact with AI in the story world. Here are the detailed descriptions for each mode:
1. Screenwriter as an Internal Character
- In the first box on the left, the screenwriter is represented as an internal character within the story world, such as playing the role of Rose.
- In this setup, the screenwriter interacts with the AI, which plays another character, like Jack.
- Through communication (indicated by two-way arrows), the screenwriter as Rose engages in dialogue and interaction with the AI as Jack.
- This means the screenwriter directly participates in the story world, experiencing and advancing the narrative as an internal character.
2. Screenwriter as an External Character
- The middle box shows the screenwriter as an external character, such as a bystander who does not directly participate in the core plot.
- The AI still plays the main role of Jack, while the screenwriter, as an external character, communicates and interacts with the AI's character.
- In this mode, there is still interaction between the screenwriter and the AI character, but the screenwriter is not a central character, participating in the story from a secondary perspective.
3. Screenwriter as an Observer
- In the third box on the right, the screenwriter is positioned as an observer, acting in their role as a screenwriter rather than a character.
- In this case, the screenwriter does not directly participate in the story but instead watches the interaction between AI characters. The chart shows the AI playing both Rose and Jack, while the screenwriter observes their dialogue.
- In this mode, the screenwriter does not intervene in the development of the story but gathers inspiration or insights from watching the interactions between the AI-controlled characters.}
 \label{actor}
 \end{figure*}

After expressing their functional expectations, all participants emphasized the need to combine AI-generated content with customizable parameters for preliminary setup. They expect to define characters through parameters such as writing style, detailed story outline, character information, backstories, reactions in specific situations, and current scenarios. Following this setup, 18 participants expressed a strong desire to enhance their portrayal of characters by engaging in dialogue with the AI ``actor.'' This feature allows screenwriters to maintain control and guide the AI to obtain the necessary information. Ideal interactions with the AI ``actor'' suggested by participants fall into three categories (Using \textit{Titanic} as an example, where Jack and Rose are the main characters and AI is playing the role of Jack, there are three categories of screenwriter engagement as illustrated in Fig. \ref{actor}.): (1) the screenwriter interacts with AI Jack as a character within the screenplay’s story world (e.g., screenwriter as Rose), (2) the screenwriter engages with AI Jack as a character outside the screenplay’s story world (e.g., screenwriter as a bystander), and (3) the screenwriter observes interactions between AI Rose and AI Jack (e.g., screenwriter as an observer). Meanwhile, some participants (\textcolor{black}{P2, P6, P9, P13, P15, N1, N4, and N5}) hoped to seamlessly switch between these methods to offer more choice and immersive experience, as characters reveal different facets to different people. As \textcolor{black}{N1} stated: “\textit{If you view a character from just one perspective, it's impossible to create a well-rounded character. Since they show different sides to different people.}”

\textbf{Screenwriter as an Internal Character.} The first category involves interacting with the AI ``actor'' as specific characters within the screenplay’s story world (\textcolor{black}{P2, P4, P5, P7, P9, P10, P11, P15, P16, N1, N2, and N4}). \textcolor{black}{P10} suggested that refining character portrayals through multiple character perspectives within the story world would lead to more precise and nuanced character development. They argued that this form of interaction could directly influence the final screenplay, showing specific behaviors. As \textcolor{black}{P16} noted: ``\textit{I feel that directly playing a character benefits my creativity. By embodying a character, I can better understand how another character would react in a given situation... it helps to think about what each character would do or say in the moment. This immersive experience is very helpful for the creative process.}''

\textbf{Screenwriter as an External Character.} The second category involves interacting with the AI ``actor'' as characters outside the screenplay’s story world, such as the screenwriter, a bystander, or a family member not depicted in the screenplay (\textcolor{black}{P2, P6, P10, P11, P12, P13, P15, N1, N3, and N4}). Since these characters are not directly tied to the screenplay, the interactions do not immediately affect the final text. The goal is to gain a deeper understanding of the character through the perspectives of additional personas imagined by the screenwriters, creating more realistic and well-rounded characters enriched by detailed backgrounds and biographies. \textcolor{black}{P2} noted, ``\textit{It's necessary to use a 'god's eye view' to see the various forms a character takes. It's very effective!}'' \textcolor{black}{P11} compared talking to AI characters to interviewing real people, saying, ``\textit{I might, as the screenwriter, directly ask the character about the plot, reading them dialogue from my script and seeing how they respond.}''


\textbf{Screenwriter as an Observer.} When screenwriters choose to observe interactions between AI ``actors'' as bystanders, they prefer the AI to generate content autonomously, providing inspiration for the screenplay, especially in terms of characters' dialogue and behaviors (\textcolor{black}{P2, P3, P4, P6, P7, P8, P9, P10, P11, P12, P13, P14, P15, P17, P18, N1, N3, N4, and N5}). \textcolor{black}{P2 and N3} noted that character interaction processes generate stories, which are fundamental to screenwriting. Meanwhile, screenwriters also want AI-generated content to incorporate more specific user settings to avoid random generation (\textcolor{black}{P3, P4, P11, and P12}). \textcolor{black}{P3} stated, ``\textit{For screenwriting, I hope it’s the interaction between those characters, not between them and me, but I want to be able to guide them.}'' \textcolor{black}{P11} added that if AI characters evolved to a level close to human self-awareness, allowing them to act freely in pre-set scenes could lead to more interesting interactions. \textcolor{black}{P4} emphasized the expectation of AI autonomy but noted that completely random behaviors could be inefficient: ``\textit{There are too many possibilities with random generation, making it challenging to use. It would be more convenient if I could control it in stages.}''


\subsection{Audience: Assisting Screenwriters in Evaluating the Value of Their Creations from the Perspective of Diverse Audience Groups, Increasing the Acceptance and Recognition of Their Work}
\textcolor{black}{P2, P4, P5, P6, P13, and N2 envisioned AI acting as an ``audience,'' offering timely, multidimensional feedback to enhance the marketability of screenplays.} As shown in Table~\ref{tab:expectations and stages}, this role could be applied in the goal \& idea, character, story structure \& plot, and screenplay text stages. Feedback during these stages could refine ideas, enhance character development for emotional resonance, and align final screenplays with market expectations.

\subsubsection{Evaluating as Mass Audience}
By simulating mass audience feedback, AI can help screenwriters adjust their creative direction to better align their work with audience expectations. \textcolor{black}{N2} said, ``\textit{AI represents the average level presented by the database it represents... You don’t need to show the book to ten different people from different cultural backgrounds; just show it to an AI.}'' This approach allows screenwriters to adjust their content promptly to meet the emotional expectations of the audience. \textcolor{black}{P2} argued that AI, through extensive data analysis, could distill ``\textit{the things that truly resonate with everyone,}'' thereby achieving broader emotional resonance. This capability would make AI an indispensable tool in the creative process, helping screenwriters craft works that touch the audience deeply. \textcolor{black}{P4} further emphasized that AI could provide ``\textit{purely neutral feedback,}'' which is particularly important for screenwriters who face interference from clients or other stakeholders during the creative process, as AI’s objectivity can offer more impartial references for screenplay revisions. \textcolor{black}{P5} stated that during the creative process, screenwriters often struggle to determine how to gain broader audience recognition, and AI’s data-driven feedback can help screenwriters approach the optimal solution. He suggested that AI should be involved in every stage of the creative process, including screenwriting, directing, and post-production, to ensure that the work maximizes its appeal to the audience. \textcolor{black}{P13} remarked, ``\textit{AI helps my work achieve both critical acclaim and commercial success... It helps us figure out what to focus on sooner and what to avoid.}'' Through AI’s feedback mechanism, screenwriters can identify the market potential of their work earlier and make targeted optimizations, thereby increasing its commercial value. 

\subsubsection{Providing Feedback as a Specific Group of Audience}
In addition to focusing on mass audiences, participants emphasized the importance of considering minority and underrepresented groups, such as individuals from specific cultural backgrounds. \textcolor{black}{P6} noted that audience preferences vary, and AI could simulate responses from diverse cultural, regional, and social backgrounds, providing feedback to help screenwriters make adjustments before releasing their work. This approach would help them avoid creative missteps caused by cultural differences or biases against minority groups.


\subsection{Expert: Offering Authoritative Evaluations and Suggestions, Guiding New Workflows Through Professional Expertise}
\textcolor{black}{P1, P4, P5, P6, P7, P8, P10, P11, P13, P15, N1, N2, and N4 expected AI to serve as an ``expert,'' offering reliable guidance and alternative solutions. As shown in Table~\ref{tab:expectations and stages}, they anticipated that the ``expert'' role could support all workflow stages, providing advice to address their limitations despite their professional training.}



\subsubsection{Providing Professional Guidance and Optimization Suggestions}
Participants \textcolor{black}{who have previously used AI in screenwriting} expect it to act as a mentor, advisor, or consultant. They anticipate that AI could identify contradictions and logical issues in the screenplay and suggest improvements, such as adding details, adjusting pacing, refining themes, modifying dialogue, revising story direction, optimizing core expressions, and enhancing aesthetic appeal (\textcolor{black}{P6, P7, P8, P10, and P11}). Additionally, participants expect AI to provide direct solutions, such as character design and plot modifications, to improve both efficiency and quality (\textcolor{black}{P5}). AI is also seen as a potential tool to assist less experienced film professionals, helping them avoid common pitfalls and enhancing the overall capabilities of the team (\textcolor{black}{P5, and P7}).

\subsubsection{Promoting New Workflows and Cultivating Multi-skilled Filmmakers}
Participants have high expectations for AI-generated image tools to foster new workflows and support multi-skilled filmmakers. \textcolor{black}{N1} stated, ``\textit{Screenwriting primarily involves writing... but imagining scenes can be exhausting... Ideally, AI should not only convert descriptions into images but also add creativity to these visuals, helping screenwriters understand the visual aspects of later production.}'' \textcolor{black}{N4} remarked, ``\textit{I prefer directing and cinematography, and I write only out of necessity... If AI could continuously provide high-quality screenplays, I would gladly use it.}'' \textcolor{black}{P4} highlighted AI’s potential to reduce costs and provide visual references for filmmakers who take on multiple roles, such as directors and cinematographers. \textcolor{black}{P15} noted that AI's visual capabilities could integrate story and visuals early, minimizing unnecessary revisions: ``\textit{If the visuals are established while the screenwriter is creating the world, nothing will be wasted, leading to greater efficiency.}''

\subsection{Executor: Fulfilling Creative Tasks According to Specified Demands, Improving Work Efficiency} \label{sec:Executor}
\textcolor{black}{Except for P1, P2, P10, P15, and N1, all other participants wanted AI to act as an ``executor,'' performing tasks based on user-defined requirements. As shown in Table~\ref{tab:expectations and stages}, they expected this role to be applied mainly in the character, story structure \& plot, and screenplay stages. With clear objectives but uncertainty about outcomes, participants viewed AI as a tool to reduce the effort of translating goals into actionable results.}




\subsubsection{Managing Continuation}
P4, P6, P8, P11, P12, P14, P17, P18, N2, N3, N4, and N5 expressed the desire for AI when acting as an ``executor,'' to better manage tasks related to the continuation and expansion of existing content. While current tools offer some capability in this area, a gap remains between participants' expectations and current standards. 

\textbf{Refining Details.} 
Participants emphasized the need for AI to ensure contextual coherence (\textcolor{black}{P4, P6, P12, N3, and N5}), as well as to enhance and polish specific details (\textcolor{black}{P8, P11, P17, P18, N2, and N5}). \textcolor{black}{N3} remarked, ``\textit{Coherence in a screenplay is crucial... it must first understand the details... then process and generate something according to the logic, not its own creation.}'' \textcolor{black}{P18} hoped AI could refine details like dialogue tone and attitude to more accurately reflect both the character's and the screenwriter’s intent.

\textbf{Generating Complete Screenplays.} In addition to managing detailed tasks, \textcolor{black}{P14, and N4} expressed a desire for AI to generate complete screenplays. They argued that even with AI advancements, high-level screenwriters would still play a key role. \textcolor{black}{P14} stated, ``\textit{Even if AI does the entire screenplay, it still needs someone with taste to filter and organize the content... A screenwriter would just have a better tool.}'' \textcolor{black}{N2} added that if AI could generate a cohesive story with a rich plot and character depth, it would be a significant achievement.

\subsubsection{Visualizing Presentation} \label{sec:Presentation}
While previous studies have addressed some challenges through visualization, participants in this study expressed specific expectations for visualizing the plot, character relationships, and emotions.

\textbf{Visualizing Plot Structures and Character Relationships.} \textcolor{black}{P3, P6, P7, P9, P12, P13, P14, P16, P18, N3, and N4} highlighted the potential of AI’s visualization function in enhancing screenwriting efficiency. \textcolor{black}{P3} suggested that AI could organize screenplays using mind maps, helping to clarify complex plot structures. \textcolor{black}{P6} added, ``\textit{When I don’t know how to advance the story, AI can generate new plot points visually and show how they connect with existing content,}'' streamlining the creation process. \textcolor{black}{P9} envisioned AI creating connections between characters within the story, generating multiple plot paths and endings.


\textbf{Visualizing Emotional Rhythms.} \textcolor{black}{P5, P7, P12, P13, P14, P17, and N3} hoped AI could abstract complex emotional cues into visual content. \textcolor{black}{P7} proposed using curves to represent emotional changes, with time on the horizontal axis and emotional fluctuations on the vertical, helping screenwriters manage emotional rhythm. He also suggested AI could display the proportional impact of events on a character’s emotions using pie charts, such as 40\% family, 30\% love, and 10\% work, which would be valuable for analysis. \textcolor{black}{N3} added that AI could use visualization to track characters’ emotional states, making the emotional narrative more coherent.



\subsection{Summary}
Participants identified four key roles for AI in screenwriting: ``actor,'' ``audience,'' ``expert,'' and ``executor.'' \textcolor{black}{As an ``actor,'' AI simulates characters, particularly in the goal \& idea, character, and dialogue stages. In the ``audience'' role, AI provides feedback during the goal \& idea, character, story structure \& plot, and screenplay text stages. As an ``expert,'' AI supports all stages by offering guidance. In the ``executor'' role, AI transforms ideas into outcomes, primarily applied in the character, story structure \& plot, and screenplay text stages. These roles underscore AI's ability to complement human creativity rather than replace it.}

\section{Future Design Opportunities} \label{sec:Opportunities}

\textcolor{black}{This section summarizes participants' needs identified in the findings and proposes design opportunities. Section 6 emphasizes the need for flexible AI systems capable of simulating diverse ``actor'' and ``audience'' roles to enable nuanced character portrayals and predictive audience feedback. Current tools are constrained by basic, static, and discrete emotional representations~\cite{assunccao2022overview, goyal2010toward}, failing to meet participants' expectations for continuous and dynamic emotional simulation. Additionally, while existing tools leverage big data for screenwriting and feedback analysis~\cite{pavel2015sceneskim, sanghrajka2017lisa, mutlu2020future}, participants highlighted the importance of managing complex character dynamics. Findings from Section~\ref{sec:Allocation} further underscore the need for AI to support a range of tasks in different stages, including goal \& idea generation, and story structure \& plot development (Table \ref{tab:Task Allocation}). Moreover, Section~\ref{sec:Presentation} and Section~\ref{sec:Simulating} emphasize the significance of immersive representations. Based on these findings, we identify two key opportunities: model development and interaction design. These provide guidance for developing future AI systems that address screenwriters' needs for emotional intelligence, role adaptability, and multithreaded outputs.}


\subsection{\textcolor{black}{Model Development}}  
\textcolor{black}{
We envision that there are two crucial steps, data collection and model training, in the lifecycle of model development can be improved based on findings from our research.}

\subsubsection{\textcolor{black}{Data Collection}}  
\textcolor{black}{Based on our findings mentioned in Section~\ref{sec:Opportunities}, building multimodal, fine-grained datasets across diverse scenarios is essential for addressing screenwriters' needs. We propose three key dimensions for acquiring data: social media feedback, high-quality works, and crowdsourced observations.}

\textbf{\textcolor{black}{Social Media Feedback.}}  
\textcolor{black}{Social computing methods analyze audience feedback derived from social media activity, fan discussions, and sentiment trends. These analyses provide valuable insights into emotional responses and narrative expectations across various demographics. By categorizing feedback based on genre preferences, emotional resonance, and storytelling trends, AI systems can better align their outputs with specific screenplay genres. For ``actor'' roles, social data offers insights into audience perceptions of character traits, relationships, and dynamics, thereby enhancing the AI's ability to generate resonant and relatable characters.}

\textbf{\textcolor{black}{High-Quality Films and Screenplays.}}  
\textcolor{black}{Analyzing high-quality films and screenplays uncovers patterns in character relationships, backstories, and narrative evolution. These insights could contribute to the creation of robust datasets that train AI systems to produce realistic character portrayals and interconnected storylines. This approach aids screenwriters by enriching the depth and complexity of character development within intricate narratives.}

\textbf{\textcolor{black}{Crowdsourced Observation Networks.}}
\textcolor{black}{Crowdsourcing, widely used in the creative domain~\cite{kim2017mechanical, kim2014ensemble, huang2020heteroglossia}, supports observation networks as a scalable method to collect diverse data for AI training. Engaging individuals from varied geographic and cultural contexts, these networks provide insights for creating realistic, contextually nuanced characters and settings. Standardized guidelines can specify data requirements and recording methods, while mobile apps facilitate multimodal uploads, such as text, images, audio, and video. Incentive mechanisms and validation processes, including peer reviews or AI checks, ensure data reliability and quality.}


\subsubsection{\textcolor{black}{Model Training}} \label{sec:Model Training}


\textcolor{black}{Section~\ref{sec:Current Positive} and Section~\ref{sec:Simulating} highlight the need to deepen AI’s understanding of contextual factors and characters' internal emotions, while Section~\ref{sec:Presentation} emphasizes narrative-driven emotional integration, focusing on plot context and character relationships. Accordingly, we provide suggestions for training models that are centered on context and characters. Meanwhile, to align with screenwriters' expectations for conflict-driven narratives~\cite{10.1145/3656650.3656688}, datasets can be structured using the theory of conflict~\cite{lawson1936theory, eisenstein2014film}, categorized into Human vs. Nature, Human vs. Society, Human vs. Human, and Human vs. Self.}

\textbf{\textcolor{black}{Contextual Emotion Modeling.}}
\textcolor{black}{Screenwriters require AI systems capable of dynamically adapting to evolving narrative contexts. Future systems should represent emotions as continuous dynamic trajectories and incorporate diverse conflict types such as Human vs. Nature and Human vs. Society.}
\textcolor{black}{To ensure that AI models can generate contextually appropriate responses, it is possible to leverage reinforcement learning from human feedback (RLHF) techniques to train attention-based language models that can effectively capture the contextual information following humans' approaches~\cite{niu2021review}.}
Additionally, multidimensional emotion modeling~\cite{gafa2023emotivita} has the potential to enable continuous dynamic simulations, facilitating coherent emotional shifts that enhance narrative depth and maintain alignment with story progression.

\textbf{\textcolor{black}{Character Emotion Modeling.}}  
\textcolor{black}{To create emotionally engaging storytelling, AI should simulate evolving character relationships and emotional arcs~\cite{smith2019simulating}. 
To achieve the target, an additional GNN model might be trained on subtle emotional cues enhances modeling for conflicts like Human vs. Human and Human vs. Self.
It can serve as a powerful method to model the character relationships~\cite{wu2020comprehensive}, with nodes as characters and edges capturing emotional or conflict-driven interactions. Incorporating the prediction results of this GNN model has the potential to better capture the dynamic and nuanced changes in characters' emotions. 
}


\textbf{\textcolor{black}{Cross-Disciplinary Collaboration.}}  
\textcolor{black}{Advancing AI's emotional intelligence requires collaboration among cognitive scientists, psychologists, screenwriters, and AI developers~\cite{assunccao2022overview, zhao2022emotion}.} 
\textcolor{black}{A Human-in-the-Loop (HITL) approach, such as RLHF or model fine-tuning with human-labeled datasets, could bridge technical and creative expertise by integrating expert feedback into AI development~\cite{wu2023toward}.}
This approach has the potential for accurate modeling of emotional nuances and adaptability to diverse narrative contexts. The iterative process enhances AI's reliability and ensures creative alignment with screenwriters' evolving needs.

\subsection{\textcolor{black}{Interaction Design}}

\textcolor{black}{We propose tailored interaction design guidelines to effectively address the needs for emotional intelligence, role adaptability, and multithreaded dynamic outputs.}

\subsubsection{For Emotional Intelligence Enhancement}
\textcolor{black}{Based on the model training approach focused on context and characters as discussed in Section~\ref{sec:Model Training}, and reflecting the scene and character emphasis in screenwriting, we propose three interaction methods.}

\textcolor{black}{\textbf{Contextual Interactions.}  
The interface should incorporate contextual factors, allowing screenwriters to shape emotional outputs based on evolving environments and stories. Features like real-time feedback, visualizations (e.g., heatmaps or evolving arcs), and contextual manipulation tools can enable adjustments to variables like settings or plot points to refine emotional trajectories. Tools such as “contextual snapshots” can help maintain narrative coherence across scene transitions~\cite{10.1145/2508244.2508251}.}

\textcolor{black}{\textbf{Character-Driven Interactions.}  
The interface should support dynamic emotional simulations, enabling the refinement of characters’ emotional responses and relationships. Tools such as timelines or radial graphs can visualize emotional evolution, while sliders allow for fine-tuning of intensity and pace. Relationship maps can simulate changes, such as conflict or reconciliation, ensuring emotional depth and alignment with character arcs~\cite{weiland2023creating}.}


\textcolor{black}{\textbf{Context-Character Interwoven Interactions.}  
To integrate context and character dynamics, the interface should align characters’ emotions with contextual elements. Real-time scenario testing can explore interactions, such as weather changes affecting emotional states. Moreover, incorporating collaborative modes could encourage contributions from screenwriters and stakeholders, providing visualizations to align with unified storytelling objectives. Additionally, as discussed in Section~\ref{sec:Multimodal}, multimodal outputs could further enhance creativity and improve communication efficiency.}

\subsubsection{For Flexibility and Role Adaptation}
\textcolor{black}{Interactive solutions like modular input and multi-agent systems provide structured and dynamic methods for seamlessly adapting ``actor'' and ``audience'' roles in collaborative screenwriting.}


\textcolor{black}{\textbf{Modular Input.}}  
\textcolor{black}{For ``actor'' roles, modular components create detailed character maps, detailing traits, backstories, relationships, and emotional states. These interfaces allow screenwriters to dynamically refine profiles, supporting seamless and flexible role transitions~\cite{ye2023mplug}. For ``audience'' roles, modular inputs define segments by demographics, preferences, or emotional metrics, integrating with ``actor'' modules to simulate audience responses to character roles, plot points, or themes.}

\textcolor{black}{\textbf{Multi-Agent Systems.}}  
\textcolor{black}{For ``actor'' roles, characters function as independent agents interacting dynamically in a shared story world~\cite{10.1145/3586183.3606763}. Screenwriters can modify interactions and preview narrative changes to inspire creativity. For ``audience'' roles, agents simulate diverse demographics, providing feedback on engagement and exploring how narrative elements spark discussions or attention, helping screenwriters craft stories that resonate with varied audiences or encourage thematic exploration.}

\subsubsection{For Multithreaded and Multimodal Generation}\label{sec:Multimodal}
\textcolor{black}{Narrative branching and immersive visualization present potential avenues for further exploring AI's capabilities in screenwriting.}

\textcolor{black}{\textbf{Intuitive Interaction with Narrative Branches.}}  
Participants highlighted the need for AI to generate and manage diverse narrative outputs, essential for exploring creative possibilities and making informed decisions. Features supporting multiple narrative branches would enable screenwriters to experiment with plot variations and assess the broader implications of changes in story elements. Real-time previews of these branches could facilitate intuitive adjustments, enhancing both the efficiency and effectiveness of the screenwriting process.

\textcolor{black}{\textbf{Immersive Visualization of Characters and Scenes.}}  
Participants suggested visualization tools are critical for transforming abstract concepts into tangible representations, and further emphasize the potential of immersive technologies, such as extended reality (XR) and 3D projection, to enhance this process. Integrating these technologies would allow AI tools to provide real-time visualizations of characters and scenes, offering insights into dynamics, spatial relationships, and emotional pacing. This capability could significantly enrich narrative development and improve the overall screenwriting experience.

\section{Discussion}


In this section, we propose aspects that future AI tools in screenwriting should focus on to enhance human-AI co-creation. Specifically, we suggest potential solutions through the lenses of agency, identity, and training. We also acknowledge the limitations of our research and outline potential directions for future studies.

\subsection{Preserving Creative Agency}

Previous research emphasizes the importance of preserving human creativity amidst technological advancements and advocates for interdisciplinary teams to promote the responsible use of AI in creative processes~\cite{suchacka2021human}. Building on this, our study suggests that future guidelines for screenwriting should specifically consider agency in the context of human-AI collaboration. In Section~\ref{sec:Allocation}, AI integration in task allocation highlights the need for human input at each stage of the workflow. In Section~\ref{sec:Engaging} and Section~\ref{sec:Executor}, where AI acts as an ``actor'' and functions as an ``executor,'' respectively, screenwriters emphasized the importance of maintaining control over the creative process and its potential impacts. They aim to ensure that AI serves as an extension of the screenwriter's creative intent rather than acting as an autonomous creator. However, as AI takes on more complex tasks, such as dialogue, structure and plot development, and even generating complete screenplays, it could challenge a screenwriter’s control over the narrative. We suggest that future AI screenwriting tool developers consider how AI tools can grant more control to screenwriters over the output, such as by providing interactive feedback loops and improving transparency in AI decision-making. Stakeholders should also think about how to maintain or regain agency by enabling screenwriters to become more adept at directing AI outputs (as discussed in Section~\ref{sec:Facilitating}).


\subsection{Redefining Authorial Identity}

\textcolor{black}{The traditional view of screenwriters as sole creators is shifting toward a collaborative human-AI model. As noted in Section 6.1.2, screenwriters are gradually becoming curators or observers of AI-generated content in their expectations, raising concerns about originality, ownership, and creative contributions (Section~\ref{sec:ethical concerns}). To address these concerns, we propose strategies to maintain authorial identity and frame this shift as an opportunity for innovation.}

\textcolor{black}{\subsubsection{Preserving Creative Contribution through Personalization}  
AI tools can enhance creative identity by aligning outputs with individual styles. For instance, tools like Stable Diffusion enable artists to embed their style into AI training~\cite{stablediffusion2022}. Similarly, screenwriters could guide AI outputs by uploading screenplays or outlines to refine tone, pacing, and structure while preserving stylistic specifics. Additionally, prior research highlights AI’s ability to merge trends with creators’ styles in short videos~\cite{10.1145/3613904.3642476}, suggesting its potential to synthesize real-world events or news with screenwriters' preferences, balancing external influences and unique styles in screenwriting.}

\textcolor{black}{\subsubsection{Maintaining Ownership and Accountability through Transparency}  
Transparency is essential for preserving authorial identity~\cite{10.1145/3613904.3641895}. We propose tracking AI’s role at different stages, such as dialogue generation or plot development, by documenting the proportion of AI versus human contributions. This systematic breakdown allows screenwriters to reflect on their processes, recognize AI’s role, and identify areas for improvement. This approach could also safeguard creative ownership, support ethical guidelines, and inform standards for attribution and intellectual property, aiding in copyright decisions based on human and AI contributions.}

\subsection{Facilitating Collaborative Training} \label{sec:Facilitating}

\textcolor{black}{Sections~\ref{sec:Allocation} and~\ref{sec:capabilities} reveal the steep learning curve of AI tools, highlighting gaps in screenwriters' understanding and training for effective integration. Screenwriters view AI as both a competitor (Section~\ref{sec:Contradictory AI as a competitor}) and a collaborator (Section~\ref{sec:Positive Potential}), emphasizing the need for balanced training. Collaboration among AI developers, educators, and screenwriters is essential for designing effective strategies for these training programs. Inspired by Kicklighter et al.'s generative AI strategies for animation education~\cite{kicklighter2024empowering}, we propose a customized training workshop program for screenwriters.}

\subsubsection{\textcolor{black}{Understanding AI Capabilities and Ethical Considerations}}
\textcolor{black}{Workshops should begin by introducing AI's strengths, limitations, and applications, guiding screenwriters in setting realistic expectations. This is especially crucial for those unfamiliar with using AI in screenwriting. Screenwriters should recognize the value AI brings, such as generating diverse ideas during brainstorming. Ethical considerations should be embedded within the training to develop technical skills, build confidence, and enhance human-AI collaboration, ultimately improving both efficiency and creativity.}

\subsubsection{\textcolor{black}{Integrating AI into Workflow}}
\textcolor{black}{Workshops offer a practical setting for screenwriters to share experiences, explore peer strategies for AI integration, and experiment with AI across different stages of the screenwriting process. Building on Kim et al.'s approach for short video producers~\cite{10.1145/3544548.3581225}, these workshops can facilitate brainstorming sessions, enabling screenwriters to leverage AI's strengths in strategically allocating tasks between AI and humans. This approach could foster technical proficiency, build confidence, and enhance collaboration efficiency.}

\subsubsection{\textcolor{black}{Promoting Awareness of Human Strengths}}
\textcolor{black}{To balance AI's roles as both a competitor and collaborator, training should emphasize unique human strengths, such as emotional depth and personal experience, which remain beyond the capabilities of current AI (Sections~\ref{sec:capabilities} and~\ref{sec:ethical concerns}). Workshops should actively encourage reflection on these unique human attributes, promoting the exploration of innovative strategies to maintain creative control while effectively utilizing AI tools.}

\subsection{Limitation and Future Work}

Although our study provides valuable insights into the integration of AI in screenwriting, it is not without limitations. One primary limitation is that all participants were from China, which may introduce cultural and regional biases in the results. Additionally, the participants’ professional backgrounds and years of screenwriting experience were not directly correlated, as all information was self-reported by the participants. Furthermore, proficiency levels in AI usage among participants varied, potentially leading to differing perspectives. However, we did not specifically compare the viewpoints of participants with different AI proficiency levels, which could be a direction for future research. As AI capabilities continue to expand, there is value in exploring how advancements in AI technology might influence screenwriters' job satisfaction, professional identity, and authorship recognition in the context of human-AI co-creation. Additionally, the development of more intuitive AI interfaces and the commercialization or open-sourcing of advanced AI tools could make these technologies more accessible to a broader range of screenwriters. By addressing these limitations and expanding the scope of research, the field can move closer to realizing the potential of AI in screenwriting, ultimately providing screenwriters with more innovative and effective tools.

\section{Conclusion}

In this study, we conducted a qualitative analysis of interviews with 23 screenwriters to explore their feedback on current practices, attitudes, and future roles of AI in screenwriting. The majority (78\%) have integrated AI into traditional workflows, particularly in the stages of story structure \& plot, screenplay text, goal \& idea generation, and dialogue. \textcolor{black}{Based on their responses, we gained deeper insights into screenwriters' attitudes toward AI integration, which influence various workflow stages and the broader industry.} We also identified four AI roles for future expectations: actor, audience, expert, and executor. \textcolor{black}{Based on these findings, we recommend that future research focus on enhancing emotional intelligence, improving role flexibility, and advancing multithreaded and multimodal generation. Furthermore, promoting human-AI co-creation should prioritize preserving creative agency and identity while facilitating collaborative training.} Overall, this study provides suggestions for future AI tools in screenwriting.

\begin{acks}
This work was partially supported by the Hong Kong RGC GRF grant 16218724 and the EdUHK-HKUST Joint Centre for Artificial Intelligence (JC\_AI) research scheme (Grant No. FB454). We extend our gratitude to Rui Sheng, Linping Yuan, Runhua Zhang, Liwenhan Xie, Xian Xu, and Yujia He for their valuable feedback and discussions on this work. We also sincerely thank all participants for their time and effort. Lastly, we greatly appreciate the reviewers for their insightful and constructive suggestions.
\end{acks}


\bibliographystyle{ACM-Reference-Format}
\bibliography{software}

\appendix

\end{document}